\documentclass[pdflatex,sn-basic]{sn-jnl}

\usepackage{graphicx}%
\usepackage{multirow}%
\usepackage{amsmath,amssymb,amsfonts}%
\usepackage{amsthm}%
\usepackage{mathrsfs}%
\usepackage[title]{appendix}%
 
\usepackage[T1]{fontenc}
\usepackage{lmodern}
\usepackage{mathrsfs}
\usepackage[font=small,labelfont=bf]{caption}
\usepackage{amsmath}
 
\usepackage{float}

\usepackage{textcomp}%
\usepackage{manyfoot}%
\usepackage{booktabs}%
\usepackage{algorithm}%
\usepackage{algorithmicx}%
\usepackage{algpseudocode}%
\usepackage{listings}%
\usepackage{hyperref}%
\usepackage{amsmath}
\usepackage{amssymb}
\usepackage{caption}
\usepackage{booktabs}
\usepackage{amsmath,amssymb,amsfonts}
\usepackage{multicol}
\usepackage{caption}
\usepackage{subcaption}
\usepackage{setspace}
\usepackage{textcomp}
\usepackage{mathtools}
\usepackage{amsmath}
\usepackage{xurl}
\usepackage{rotating}
\usepackage{lscape}
\usepackage{longtable}
\usepackage{hyperref}
\usepackage{natbib,stfloats}
\usepackage{mathrsfs}
\usepackage{hyperref}
\usepackage{lscape}
\usepackage{listings}
\usepackage[table,xcdraw]{xcolor}
\usepackage{colortbl}

\makeatletter
\newcommand\subsubsubsection{\@startsection{paragraph}{4}{\z@}%
  {3.25ex \@plus1ex \@minus.2ex}%
  {-1em}%
  {\normalfont\normalsize\bfseries}}
\makeatother

\begin{document}

\title[Benchmarking 46 Polygenic Risk Score Tools]{A harmonized benchmarking framework for implementation-aware evaluation of 46 polygenic risk score tools across binary and continuous phenotypes}


\author[1,2]{\fnm{Muhammad} \sur{Muneeb}}\email{m.muneeb@uq.edu.au}
\author*[1,2]{\fnm{David}  \sur{B. Ascher}}\email{d.ascher@uq.edu.au}

\affil*[1]{\orgdiv{School of Chemistry and Molecular Biology}, \orgname{The University of Queensland}, \orgaddress{\street{Queen Street}, \postcode{4067}, \state{Queensland}, \country{Australia}}}

\affil[2]{\orgdiv{Computational Biology and Clinical Informatics}, \orgname{Baker Heart and Diabetes Institute}, \orgaddress{\street{Commercial Road}, \postcode{3004}, \state{Victoria}, \country{Australia}}}
 
\abstract{Polygenic risk score (PRS) tools differ substantially in statistical assumptions, input requirements, and implementation complexity, making direct comparison difficult. We developed a harmonized, implementation-aware benchmarking framework to evaluate 46 PRS tools across seven binary UK Biobank phenotypes and one continuous trait under three model configurations: null, PRS-only, and PRS plus covariates. The framework integrates standardized preprocessing, tool-specific execution, hyperparameter exploration, and unified downstream evaluation using five-fold cross-validation on high-performance computing infrastructure. In addition to predictive performance, we assessed runtime, memory use, input dependencies, and failure modes. A Friedman test across 40 phenotype--fold combinations confirmed significant differences in tool rankings ($\chi^2 = 102.29$, $p = 2.57 \times 10^{-11}$), with no single method universally optimal. These findings provide a reproducible framework for comparative PRS evaluation and demonstrate that tool performance is shaped not only by statistical methodology but also by phenotype architecture, preprocessing choices, covariate structure, computational demands, software robustness, and practical implementation constraints. }

\keywords{polygenic risk scores, benchmarking framework, implementation-aware evaluation, genetic risk prediction, cross-validation, bioinformatics}
 


\maketitle

\newpage
\section*{Introduction}
PRS quantify an individual’s genetic predisposition to complex traits, diseases, and related outcomes by aggregating effect-size estimates from genome-wide association studies (GWAS) across many single-nucleotide polymorphisms (SNPs)~\cite{Choi2020,Lewis2020,RN8133,Nakase2024}. PRS construction can involve multiple data sources, including GWAS summary statistics~\cite{Nakase2024}, genotype data~\cite{Hui2023}, covariates such as age, sex, and environmental factors~\cite{Hui2023}, reference panels~\cite{Loh2023}, linkage disequilibrium (LD) information~\cite{Vilhjalmsson2015}, and, in some pipelines, GWAS from related traits or populations~\cite{Zhuang2024}. Although the underlying PRS formulation is conceptually simple, accurate score construction depends on how SNP effects ($\beta$) or posterior quantities are estimated under different modelling assumptions~\cite{ZHU20201557}. This has led to a broad range of PRS methods, including linear mixed models (LMMs), generalized linear mixed models for non-normal traits, Bayesian variable-selection and shrinkage approaches such as Bayesian variable selection regression (BVSR) and Bayesian sparse linear mixed models (BSLMM), sparsity-based models, and cross-trait or multi-ancestry extensions~\cite{Loh2023,ZHU20201557}.

As GWAS sample sizes and the availability of public summary statistics have increased, the number of computational tools for PRS construction has also increased~\cite{Choi2020,Lewis2020}. These tools differ in statistical assumptions, treatment of LD~\cite{Vilhjalmsson2015}, use of functional annotation~\cite{Zhuang2024}, reliance on individual-level versus summary-level data~\cite{Hui2023}, computational implementation~\cite{ZHU20201557}, and the type of phenotype or study design for which they are best suited~\cite{Ma2021,Zhang_2021}. As a result, researchers face a broad and heterogeneous PRS software landscape when selecting an appropriate method for a given analysis.

Several comparative studies have evaluated subsets of PRS methods and shown that performance depends strongly on phenotype architecture, training data characteristics, sample composition, and modeling choices~\cite{Ma2021,Zhang_2021}. However, direct comparison across the expanding PRS ecosystem remains difficult. Existing benchmarks often evaluate only a limited number of tools, use heterogeneous preprocessing and validation strategies, or assess performance in narrowly defined settings~\cite{Ma2021,Zhang_2021}. Most studies emphasize predictive performance alone and pay limited attention to practical implementation factors, including installation complexity, software dependencies, input-format requirements, runtime, memory usage, reference-panel needs, and failure behavior under real-data constraints. These operational characteristics can strongly influence whether a method is practical for large-scale analyses, particularly in high-performance computing (HPC) environments and multi-phenotype benchmarking studies.

Another challenge is that comparative performance may reflect not only the PRS method itself, but also the analytical setting in which the score is evaluated. Covariate structure, phenotype type, preprocessing choices, and hyperparameter selection can all influence downstream prediction~\cite{Hui2023,Ma2021,Zhang_2021}. Benchmarks that report only a single final performance measure may therefore obscure whether observed differences arise from the PRS model itself or from the surrounding modelling pipeline. A fair comparison requires standardized preprocessing, consistent validation, and explicit separation of null, PRS-only, and PRS-plus-covariate models.

\begin{figure}[!ht]
\centering
\includegraphics[width=0.80\textwidth]{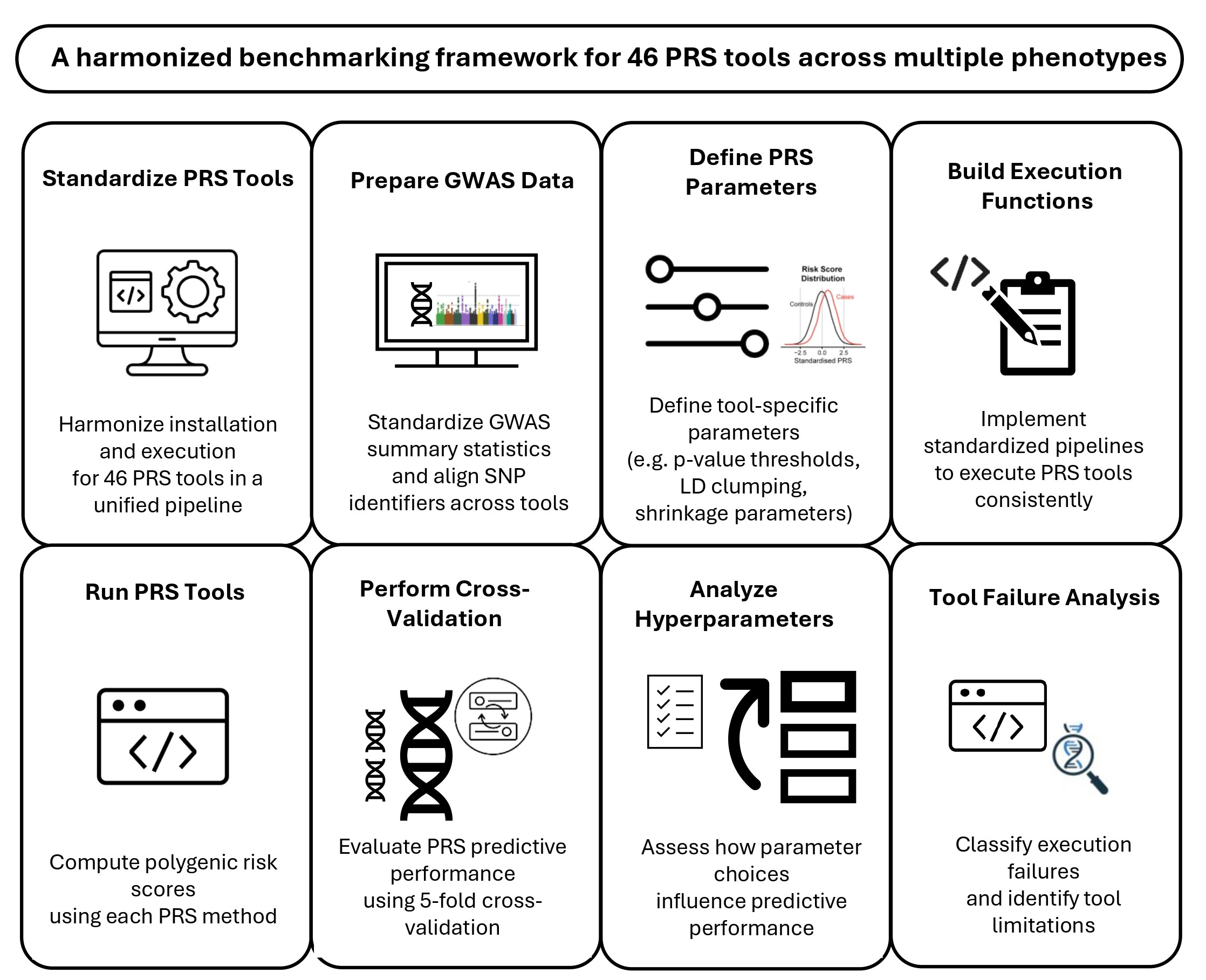}
\caption{\textbf{Harmonized benchmarking framework used to evaluate 46 PRS tools across eight phenotypes.} The framework standardizes installation, data preparation, hyperparameter definition, and tool execution, followed by cross-validation, performance evaluation, hyperparameter sensitivity analysis, and a structured analysis of the tool's results.}
\label{NotebookWorkflowFinal}
\end{figure}

To address these gaps, we developed a harmonized benchmarking framework for implementation-aware comparative evaluation of PRS tools under standardized preprocessing, execution, and validation conditions. Using this framework, we benchmarked 46 PRS tools across seven binary phenotypes from the UK Biobank and one continuous phenotype (Height). The framework integrates data transformation, tool-specific execution on HPC systems, five-fold cross-validation, hyperparameter exploration, and unified downstream evaluation. Each phenotype was evaluated under three model configurations: a null model, a pure PRS model, and a full model combining PRS with covariates. For binary phenotypes, predictive performance was assessed using the area under the receiver operating characteristic curve (AUC), whereas for Height, performance was assessed using explained variance ($R^2$), as summarized in Figure~\ref{NotebookWorkflowFinal}.

This study makes three main contributions. First, it establishes a reproducible benchmarking framework that standardizes data preparation, execution, and evaluation across a large and heterogeneous set of PRS tools. Second, it provides a comparative analysis across phenotypes, showing that PRS performance varies across traits and analytical settings, with no single method consistently outperforming all others. Third, it documents implementation-aware aspects of PRS benchmarking, including runtime, memory usage, installation requirements, input dependencies, and structured failure modes. Together, these features position the study as a practical resource for transparent and reproducible PRS benchmarking rather than as a definitive ranking of universally superior tools. The full framework, code, and documentation are publicly available at \url{https://muhammadmuneeb007.github.io/PRSTools/Introduction.html}.


\section*{Methods}
\label{Methodology}

\subsection*{Dataset}
We benchmarked 46 PRS tools across eight phenotypes: seven binary phenotypes from the UK Biobank formed the primary multi-phenotype benchmark, and one continuous phenotype (Height) was included as a well-characterised continuous-trait case study using an independent publicly available tutorial dataset~\cite{Choi2020}. The two datasets differ in cohort provenance, sample size, and covariate structure, and results for Height should therefore be interpreted as a complementary case study rather than as a directly comparable eighth arm of the binary benchmark. For the binary phenotypes, genotype data were extracted for 733 participants, and the corresponding GWAS summary statistics were obtained from the GWAS Catalog (\url{https://www.ebi.ac.uk/gwas/}). The genotype dataset included individuals with reported comorbid conditions, including hypertension, asthma, depression, osteoarthritis, high cholesterol, irritable bowel syndrome, hypothyroidism, hay fever, migraine, and gastric reflux, together with 135 nuclear magnetic resonance (NMR) metabolomic biomarkers. These biomarkers and comorbid conditions were used as covariates in downstream predictive models. For the continuous Height phenotype, we used the GWAS and genotype data provided in the PRS tutorial dataset described by ~\cite{Choi2020}, available at \url{https://choishingwan.github.io/PRS-Tutorial/base/}. For this dataset, age and sex were used as covariates.


\subsection*{Quality control and harmonization}
GWAS summary statistics for all phenotypes were obtained from the GWAS Catalog summary statistics resource (\url{https://www.ebi.ac.uk/gwas/downloads/summary-statistics}), and the corresponding study-specific summary files are listed in Supplementary Table~1 (\textbf{GWAS Information}). To enable consistent benchmarking across tools, GWAS summary statistics and genotype data were harmonized before tool-specific execution. GWAS files were standardized using GWASPokerforPRS (\url{https://github.com/MuhammadMuneeb007/GWASPokerforPRS}) to retain the fields required for PRS construction, including chromosome, base-pair position, SNP identifier, p-value, effect size or odds ratio, reference allele, alternative allele, and minor allele frequency. Genome builds were matched between the GWAS and genotype datasets before downstream analysis. 

Quality control was then applied to both GWAS and genotype data. For GWAS summary statistics, SNPs were retained if the minor allele frequency (MAF) was greater than 0.01 and the imputation INFO score was greater than 0.8. Ambiguous SNPs and SNPs with missing alleles, including complementary A/T and C/G variants, were removed to reduce strand-alignment errors. For genotype data, we applied filters of MAF $> 0.01$, Hardy--Weinberg equilibrium threshold $p > 1 \times 10^{-6}$, genotype missingness $< 0.1$, individual missingness $< 0.1$, and relatedness cutoff $< 0.125$. After quality control, the final genotype dataset for the seven binary phenotypes contained approximately 600{,}000 SNPs and 152 covariates, whereas the Height dataset contained 551{,}892 SNPs and two covariates.


\subsection*{Benchmarking workflow/execution framework}

We used a harmonized benchmarking framework in which common preprocessing steps were standardized while allowing tool-specific execution paths. PRS tools differed substantially in their assumptions, accepted inputs, software dependencies, and computational requirements. To support fair comparison, each tool was installed, configured, and executed under a standardized workflow that included data preparation, parameter definition, PRS generation, downstream model fitting, and fold-wise performance measurement.

Following preprocessing, genotype and GWAS data were reformatted to satisfy tool-specific input requirements. Some tools required PLINK binary files, whereas others required dosage files or exact SNP correspondence between GWAS and genotype datasets. Based on input structure, tools were broadly grouped into four categories: (i) methods using GWAS summary statistics only, (ii) methods using both GWAS summary statistics and genotype data, (iii) methods using genotype data only, and (iv) cross-trait or multi-ancestry methods using GWAS from multiple populations or diseases. Some methods, such as NPS and CTPR, imposed strict input requirements, including a genotype rate equal to 1 and exact SNP matching between the GWAS and genotype files. Detailed information on tool-specific input requirements, including GWAS data, genotype data, covariates, and reference panels, is provided in Supplementary Table~2 (\textbf{PRS tool characteristics}). Installation, formatting, and execution details for each tool are available in the online documentation at \url{https://muhammadmuneeb007.github.io/PRSTools/Introduction.html}.

\subsection*{Cross-validation and hyperparameter search}

All PRS tools were evaluated within a five-fold cross-validation framework. For each phenotype, the data were split into training and test sets. Tool fitting, hyperparameter exploration, and model selection were performed using the training data, and the resulting SNP weights or effect sizes were then applied to the corresponding test set. This design was used to reduce information leakage and ensure that reported performance reflected out-of-sample evaluation.

Hyperparameters explored in benchmarking included clumping, pruning, principal component adjustment, p-value thresholds, reference-panel choices, and method-specific options such as heritability estimation. For clumping, we used a p-value threshold of 1, an LD threshold of 0.1, and a distance threshold of 200. For pruning, we used a window size of 200, a step size of 50, and an LD threshold of 0.25. PRS were evaluated across 20 logarithmically spaced p-value thresholds ranging from $10^{-10}$ to 1. Additional tool-specific hyperparameters, including alternative reference panels, alternative heritability estimators, and different clumping or pruning settings, were explored where relevant.

\subsection*{Model configurations}

Each phenotype was evaluated under three model configurations:

\begin{enumerate}
    \item \textbf{Null model:} covariates and principal components (PCs) only;
    \item \textbf{PRS-only model:} PRS only;
    \item \textbf{Full model:} PRS with covariates and PCs.
\end{enumerate}

The inclusion of 135 NMR metabolomic biomarkers and comorbid conditions as covariates in the binary phenotype models warrants explicit consideration when interpreting the benchmarking results. These covariates were included because they reflect clinically relevant information available in the UK Biobank dataset and constitute the realistic prediction context in which PRS tools may be deployed. However, their inclusion means that the full model evaluates the incremental predictive value of the PRS given an already information-rich covariate structure, rather than evaluating the PRS in isolation.

To address this, results are reported separately for three model configurations---null, PRS-only, and full---allowing the contribution of the PRS to be distinguished from that of the covariate set. The null and PRS-only models, in particular, provide a more covariate-independent basis for comparing tool performance, and readers should refer to these when evaluating the PRS method itself rather than the full prediction pipeline. Nonetheless, the relative ranking of tools in the full model may partly reflect interactions between specific PRS methods and the covariate structure, rather than differences in the underlying genetic modelling strategy alone. This is an inherent limitation of any benchmarking study conducted in a rich covariate setting and should be considered when generalising the results to settings with fewer covariates.

\subsection*{Performance evaluation}
For each trained model, PRS values were calculated by applying estimated SNP effect sizes to the test genotype data using PLINK \texttt{----score}~\cite{Purcell_2007,Choi2020}. PRS was computed as

\begin{equation}
PRS = \sum_{i=1}^{n} \left( \beta_i \times G_i \right),
\label{Plinkequation}
\end{equation}

where $\beta_i$ denotes the estimated effect size of the $i$-th SNP and $G_i \in \{0,1,2\}$ denotes the genotype count for that SNP. The total number of SNPs included in scoring is represented by $n$.

For binary phenotypes, predictive performance was evaluated using logistic regression and quantified using the area under the receiver operating characteristic curve (AUC)~\cite{Sperandei2014}. For the continuous Height phenotype, predictive performance was evaluated using ordinary least squares regression and quantified using explained variance ($R^2$)~\cite{Zdaniuk2014}. Performance was recorded on both training and test sets for all retained p-value thresholds and tool-specific hyperparameter combinations across the five folds.

\subsection*{Hyperparameter summarization rule}
To summarise the large hyperparameter search space consistently, we applied a predefined $\delta$-constrained selection rule to each tool and phenotype. For binary phenotypes, $\delta$ was set to 0.05 AUC, and for Height, $\delta$ was set to 0.03 in explained variance ($R^2$). Among candidate configurations with an absolute train--test performance difference below $\delta$, we selected the configuration with the highest combined training and test performance. If no configuration satisfied this criterion, we selected the configuration with the smallest train--test gap.

This selection rule was used as a practical summarisation strategy for a large hyperparameter search space and should not be interpreted as a substitute for fully nested cross-validation. Because exhaustive nested cross-validation was not computationally feasible across all tools, phenotypes, folds, and hyperparameter combinations, we applied a uniform $\delta$-constrained rule to favour configurations with smaller train--test discrepancies and more stable out-of-sample behaviour. The same $\delta$ thresholds were applied across all tools to preserve comparability. However, because test-set performance contributed to configuration summarisation, the selected configuration should be interpreted as a stability-oriented summary of the explored search space rather than as an independently validated optimum. The reported full-model results, therefore, correspond to the selected stable configuration, not necessarily to the configuration with the highest test performance. 

To evaluate whether the $\delta$-constrained selection rule materially influenced the benchmarking conclusions, we compared the main rule against three alternatives applied to all 46 PRS tools across all eight phenotypes and five cross-validation folds: (A) a training-only rule selecting the configuration with the highest training performance and no reference to test performance, representing a scenario with no overfitting control; (B) a stricter stability criterion ($\delta \times 0.5$); and (C) a more lenient criterion ($\delta \times 2.0$). Tools were ranked within each of the 40 phenotype--fold datasets under each rule and compared using Spearman rank correlation and the Friedman test.

\subsection*{Statistical comparison of PRS tools}

To assess whether PRS tool performance differed significantly across phenotypes, we performed a global ranking-based comparison using fold-level test performance from the full model. For each phenotype and fold, tools were ranked according to test performance, with rank 1 assigned to the best-performing tool. These rankings were then compared across phenotype--fold combinations using the Friedman test, a non-parametric repeated-measures method for comparing multiple algorithms across multiple datasets. When the Friedman test indicated significant differences, post hoc pairwise comparisons were performed using the Nemenyi procedure. Multiple testing correction was applied using the Benjamini--Hochberg false discovery rate (FDR) method. Average ranks across all phenotype--fold combinations were used to identify tools that performed consistently well.

\section*{Results}
\subsection*{Predictive performance varies across PRS tools and phenotypes}
We evaluated 46 PRS tools across eight phenotypes and summarized performance using the full predictive model, defined as PRS combined with covariates and principal components. For binary phenotypes, predictive performance was measured using AUC, whereas for Height, performance was measured using $R^2$ (Table~\ref{TableBestFullModelSorted}). 
\begin{table*}[!ht]
\centering
\resizebox{\textwidth}{!}{%
\begin{tabular}{|l|l|l|l|l|l|l|l|l|}
\hline
\textbf{Tool} & \textbf{Height} & \textbf{Asthma} & \textbf{Depression} & \textbf{Gastro-Reflux} & \textbf{High Cholesterol} & \textbf{Hypothyroidism} & \textbf{IBS} & \textbf{Migraine} \\
\hline
AnnoPred \cite{Hu_2017} & \cellcolor[HTML]{B2D580}0.2791 & \cellcolor[HTML]{FEE182}0.5723 & \cellcolor[HTML]{F4E884}0.5864 & NA& \cellcolor[HTML]{FEE883}0.8538 & NA& \cellcolor[HTML]{FEE081}0.5981 & \cellcolor[HTML]{FCC07B}0.5367 \\
\hline
BOLT-LMM \cite{Loh_2015} & \cellcolor[HTML]{FCBF7B}0.1354 & \cellcolor[HTML]{FBAE78}0.5328 & \cellcolor[HTML]{F8696B}0.5059 & \cellcolor[HTML]{63BE7B}0.7528 & \cellcolor[HTML]{D3DF82}0.8805 & \cellcolor[HTML]{6EC17C}0.6665 & \cellcolor[HTML]{F3E884}0.6097 & \cellcolor[HTML]{AAD380}0.5906 \\
\hline
C+T & \cellcolor[HTML]{9ACE7F}0.3016 & \cellcolor[HTML]{E3E383}0.5885 & \cellcolor[HTML]{FDD880}0.5696 & \cellcolor[HTML]{F9EA84}0.7092 & \cellcolor[HTML]{FEEB84}0.8607 & \cellcolor[HTML]{FCC17C}0.5824 & \cellcolor[HTML]{D4DF82}0.6248 & \cellcolor[HTML]{F2E884}0.5515 \\
\hline
CTPR \cite{Chung_2019} & \cellcolor[HTML]{FEE182}0.1916 & \cellcolor[HTML]{E5E483}0.5880 & \cellcolor[HTML]{FCB479}0.5489 & \cellcolor[HTML]{FEDC81}0.6989 & \cellcolor[HTML]{F2E884}0.8663 & \cellcolor[HTML]{9BCF7F}0.6459 & \cellcolor[HTML]{FEDD81}0.5966 & \cellcolor[HTML]{FEEA83}0.5439 \\
\hline
CTSLEB \cite{Chung_2021} & \cellcolor[HTML]{FDC97D}0.1514 & \cellcolor[HTML]{FBAE78}0.5326 & \cellcolor[HTML]{F0E784}0.5884 & \cellcolor[HTML]{FCBD7B}0.6819 & \cellcolor[HTML]{ECE683}0.8692 & \cellcolor[HTML]{B8D780}0.6330 & \cellcolor[HTML]{FED980}0.5946 & \cellcolor[HTML]{F8696B}0.5219 \\
\hline
DBSLMM \cite{Yang_2020} & \cellcolor[HTML]{FCBB7A}0.1283 & \cellcolor[HTML]{BDD881}0.6008 & \cellcolor[HTML]{D1DE82}0.6054 & \cellcolor[HTML]{FEDE81}0.7002 & \cellcolor[HTML]{FCEA84}0.8618 & \cellcolor[HTML]{FEE983}0.6002 & \cellcolor[HTML]{FEE282}0.5992 & \cellcolor[HTML]{FFEB84}0.5440 \\
\hline
EBPRS \cite{Song_2020} & \cellcolor[HTML]{FFEB84}0.2076 & \cellcolor[HTML]{FEE783}0.5770 & \cellcolor[HTML]{FEEA83}0.5798 & \cellcolor[HTML]{FBA977}0.6702 & \cellcolor[HTML]{FEE883}0.8534 & \cellcolor[HTML]{FEEA83}0.6006 & \cellcolor[HTML]{FDC67C}0.5839 & \cellcolor[HTML]{F0E784}0.5523 \\
\hline
GCTA \cite{Yang_2013} & \cellcolor[HTML]{BCD881}0.2703 & \cellcolor[HTML]{FEE282}0.5727 & \cellcolor[HTML]{E6E483}0.5940 & \cellcolor[HTML]{F4E884}0.7107 & \cellcolor[HTML]{F8696B}0.5036 & \cellcolor[HTML]{D8E082}0.6186 & \cellcolor[HTML]{E4E383}0.6171 & \cellcolor[HTML]{FCBA7A}0.5357 \\
\hline
GCTB-SBayesR \cite{Zheng_2024} & \cellcolor[HTML]{FEE182}0.1907 & \cellcolor[HTML]{FEE783}0.5763 & \cellcolor[HTML]{FDD880}0.5696 & \cellcolor[HTML]{FDD880}0.6968 & \cellcolor[HTML]{FEEA83}0.8590 & \cellcolor[HTML]{FDD880}0.5924 & \cellcolor[HTML]{FDD67F}0.5925 & \cellcolor[HTML]{FDD17F}0.5397 \\
\hline
GCTB-SBayesRC \cite{Lloyd_Jones_2019} & \cellcolor[HTML]{FDCF7E}0.1609 & \cellcolor[HTML]{FEDE81}0.5701 & \cellcolor[HTML]{EAE583}0.5916 & \cellcolor[HTML]{FEE983}0.7062 & \cellcolor[HTML]{EDE683}0.8686 & \cellcolor[HTML]{FCC17C}0.5822 & \cellcolor[HTML]{F0E784}0.6113 & \cellcolor[HTML]{FDD47F}0.5401 \\
\hline
GEMMA-BSLMM \cite{Zhou_2012} & \cellcolor[HTML]{FCB579}0.1187 & \cellcolor[HTML]{FDD17F}0.5596 & \cellcolor[HTML]{FFEB84}0.5803 & \cellcolor[HTML]{83C87D}0.7435 & \cellcolor[HTML]{FFEB84}0.8603 & \cellcolor[HTML]{FA9072}0.5600 & \cellcolor[HTML]{FEDD81}0.5963 & \cellcolor[HTML]{F9EA84}0.5473 \\
\hline
GEMMA-LM \cite{Zhou_2012} & \cellcolor[HTML]{F8716C}0.0044 & \cellcolor[HTML]{FBA075}0.5223 & \cellcolor[HTML]{C8DB81}0.6102 & \cellcolor[HTML]{A1D07F}0.7349 & \cellcolor[HTML]{A9D27F}0.8999 & \cellcolor[HTML]{E8E583}0.6113 & \cellcolor[HTML]{F0E784}0.6114 & \cellcolor[HTML]{66BF7C}0.6275 \\
\hline
GEMMA-LMM \cite{Zhou_2012} & \cellcolor[HTML]{F8696B}-0.0089 & \cellcolor[HTML]{CCDD82}0.5959 & \cellcolor[HTML]{E5E483}0.5945 & \cellcolor[HTML]{B3D580}0.7296 & \cellcolor[HTML]{D2DE82}0.8811 & \cellcolor[HTML]{FCBC7B}0.5801 & \cellcolor[HTML]{76C47D}0.6696 & \cellcolor[HTML]{64BF7C}0.6289 \\
\hline
HAIL \cite{Zhao_2024} & \cellcolor[HTML]{FDC97D}0.1504 & \cellcolor[HTML]{F8696B}0.4795 & \cellcolor[HTML]{E2E383}0.5962 & \cellcolor[HTML]{FCC17C}0.6842 & \cellcolor[HTML]{C3DA81}0.8878 & \cellcolor[HTML]{EEE683}0.6089 & \cellcolor[HTML]{FDD780}0.5930 & \cellcolor[HTML]{FBB178}0.5342 \\
\hline
JAMPred \cite{Newcombe_2019} & \cellcolor[HTML]{FA9F75}0.0807 & \cellcolor[HTML]{FBB078}0.5343 & \cellcolor[HTML]{FCC17C}0.5568 & NA& \cellcolor[HTML]{FEE983}0.8552 & NA& \cellcolor[HTML]{FBA376}0.5649 & \cellcolor[HTML]{E9E583}0.5564 \\
\hline
LDAK-GenotypeData & \cellcolor[HTML]{F98670}0.0406 & \cellcolor[HTML]{DDE182}0.5906 & \cellcolor[HTML]{FEE182}0.5748 & \cellcolor[HTML]{FEDE81}0.7002 & \cellcolor[HTML]{FEEA83}0.8598 & \cellcolor[HTML]{FCBB7A}0.5793 & \cellcolor[HTML]{E9E583}0.6145 & \cellcolor[HTML]{FFEB84}0.5440 \\
\hline
LDAK-GWAS \cite{Zhang_2021} & \cellcolor[HTML]{63BE7B}0.3531 & \cellcolor[HTML]{B6D680}0.6031 & \cellcolor[HTML]{63BE7B}0.6637 & \cellcolor[HTML]{DFE283}0.7168 & \cellcolor[HTML]{FEEB84}0.8606 & \cellcolor[HTML]{F7E984}0.6048 & \cellcolor[HTML]{FEDB81}0.5955 & \cellcolor[HTML]{FDEB84}0.5452 \\
\hline
Lassosum \cite{Mak_2017} & \cellcolor[HTML]{A1D07F}0.2958 & \cellcolor[HTML]{7BC57D}0.6221 & \cellcolor[HTML]{A6D27F}0.6280 & \cellcolor[HTML]{FFEB84}0.7073 & \cellcolor[HTML]{FBEA84}0.8623 & \cellcolor[HTML]{F9EA84}0.6037 & \cellcolor[HTML]{FCBB7A}0.5781 & \cellcolor[HTML]{FFEB84}0.5444 \\
\hline
LDpred-2-Auto \cite{Priv__2020} & \cellcolor[HTML]{CBDC81}0.2562 & \cellcolor[HTML]{8FCB7E}0.6154 & \cellcolor[HTML]{F7E984}0.5851 & \cellcolor[HTML]{FDEB84}0.7079 & \cellcolor[HTML]{FEEA83}0.8599 & \cellcolor[HTML]{FDCD7E}0.5877 & \cellcolor[HTML]{F9EA84}0.6070 & NA\\
\hline
LDpred-2-Grid \cite{Priv__2020} & \cellcolor[HTML]{B3D580}0.2788 & \cellcolor[HTML]{63BE7B}0.6295 & \cellcolor[HTML]{B4D680}0.6204 & \cellcolor[HTML]{F6E984}0.7100 & \cellcolor[HTML]{FFEB84}0.8602 & \cellcolor[HTML]{FEE783}0.5993 & \cellcolor[HTML]{FCEB84}0.6054 & NA\\
\hline
LDpred-2-Inf \cite{Priv__2020} & \cellcolor[HTML]{A4D17F}0.2929 & \cellcolor[HTML]{FDCE7E}0.5575 & \cellcolor[HTML]{F6E984}0.5853 & \cellcolor[HTML]{C8DC81}0.7234 & \cellcolor[HTML]{FEEA83}0.8591 & \cellcolor[HTML]{FBA676}0.5699 & \cellcolor[HTML]{D9E082}0.6224 & NA\\
\hline
LDpred-2-Lassosum2 & \cellcolor[HTML]{C9DC81}0.2575 & \cellcolor[HTML]{7BC57D}0.6219 & \cellcolor[HTML]{92CC7E}0.6389 & \cellcolor[HTML]{FFEB84}0.7073 & \cellcolor[HTML]{FFEB84}0.8602 & \cellcolor[HTML]{C3DA81}0.6283 & \cellcolor[HTML]{FEEB84}0.6044 & \cellcolor[HTML]{F9EA84}0.5477 \\
\hline
LDpred-fast \cite{Reales_2021} & \cellcolor[HTML]{6DC17C}0.3442 & \cellcolor[HTML]{C2DA81}0.5991 & \cellcolor[HTML]{CBDC81}0.6085 & \cellcolor[HTML]{F8696B}0.6342 & NA& \cellcolor[HTML]{BDD881}0.6308 & \cellcolor[HTML]{FDCB7D}0.5867 & \cellcolor[HTML]{FFEB84}0.5444 \\
\hline
LDpred-funct \cite{M_rquez_Luna_2021} & \cellcolor[HTML]{F0E784}0.2217 & \cellcolor[HTML]{BCD881}0.6009 & \cellcolor[HTML]{FDCB7D}0.5621 & NA& \cellcolor[HTML]{FEE883}0.8546 & NA& \cellcolor[HTML]{FDC67C}0.5839 & \cellcolor[HTML]{FFEB84}0.5440 \\
\hline
LDpred-gibbs \cite{Reales_2021} & \cellcolor[HTML]{E3E383}0.2331 & \cellcolor[HTML]{FCB679}0.5390 & \cellcolor[HTML]{FBAE78}0.5456 & \cellcolor[HTML]{CDDD82}0.7220 & \cellcolor[HTML]{FEE983}0.8552 & \cellcolor[HTML]{9DCF7F}0.6453 & \cellcolor[HTML]{F8696B}0.5335 & \cellcolor[HTML]{FEE983}0.5437 \\
\hline
LDpred-inf \cite{Reales_2021} & \cellcolor[HTML]{8ACA7E}0.3166 & \cellcolor[HTML]{F5E984}0.5826 & \cellcolor[HTML]{FCBC7A}0.5535 & \cellcolor[HTML]{F4E884}0.7106 & \cellcolor[HTML]{FEE983}0.8553 & \cellcolor[HTML]{63BE7B}0.6710 & \cellcolor[HTML]{FFEB84}0.6038 & \cellcolor[HTML]{FEEA83}0.5439 \\
\hline
LDpred-p+t \cite{Reales_2021} & \cellcolor[HTML]{8ACA7E}0.3167 & \cellcolor[HTML]{7AC57D}0.6222 & \cellcolor[HTML]{DDE283}0.5987 & \cellcolor[HTML]{76C47D}0.7475 & \cellcolor[HTML]{FEEA83}0.8574 & \cellcolor[HTML]{F8696B}0.5423 & \cellcolor[HTML]{E7E483}0.6156 & \cellcolor[HTML]{FBEA84}0.5463 \\
\hline
MTG2 \cite{Lee_2016} & \cellcolor[HTML]{FA9272}0.0601 & \cellcolor[HTML]{FBB078}0.5348 & \cellcolor[HTML]{FDC67C}0.5592 & NA& \cellcolor[HTML]{7FC67D}0.9193 & \cellcolor[HTML]{A4D17F}0.6418 & \cellcolor[HTML]{6EC27C}0.6734 & \cellcolor[HTML]{71C27C}0.6219 \\
\hline
NPS \cite{Chun_2020} & \cellcolor[HTML]{FEDB80}0.1808 & \cellcolor[HTML]{D3DF82}0.5935 & \cellcolor[HTML]{FCB579}0.5494 & NA& NA& NA& \cellcolor[HTML]{D8E082}0.6226 & \cellcolor[HTML]{FCBB7A}0.5360 \\
\hline
PANPRS \cite{Chen_2020} & \cellcolor[HTML]{FEDA80}0.1802 & \cellcolor[HTML]{B5D680}0.6034 & \cellcolor[HTML]{B2D580}0.6218 & \cellcolor[HTML]{D5DF82}0.7198 & \cellcolor[HTML]{FEE983}0.8558 & \cellcolor[HTML]{B3D580}0.6354 & \cellcolor[HTML]{FEE382}0.5995 & \cellcolor[HTML]{E4E483}0.5589 \\
\hline
PleioPred \cite{Weissbrod_2022} & \cellcolor[HTML]{C0D981}0.2666 & \cellcolor[HTML]{BAD881}0.6016 & NA& NA& NA& NA& NA& NA\\
\hline
PLINK \cite{Purcell_2007} & \cellcolor[HTML]{94CD7E}0.3072 & \cellcolor[HTML]{FEE582}0.5751 & \cellcolor[HTML]{FA9A74}0.5344 & \cellcolor[HTML]{FEE783}0.7056 & \cellcolor[HTML]{FEE883}0.8536 & \cellcolor[HTML]{FED980}0.5931 & \cellcolor[HTML]{F8746D}0.5397 & \cellcolor[HTML]{FDEB84}0.5452 \\
\hline
PolyPred \cite{Hu_2017} & \cellcolor[HTML]{FCBA7A}0.1255 & \cellcolor[HTML]{FCC37C}0.5491 & \cellcolor[HTML]{98CE7F}0.6357 & NA& \cellcolor[HTML]{FCEA84}0.8617 & NA& \cellcolor[HTML]{A5D17F}0.6473 & \cellcolor[HTML]{B8D780}0.5830 \\
\hline
PRSbils & \cellcolor[HTML]{F0E784}0.2211 & \cellcolor[HTML]{FCB379}0.5367 & \cellcolor[HTML]{FAEA84}0.5835 & \cellcolor[HTML]{FCBF7B}0.6829 & \cellcolor[HTML]{FEEA83}0.8595 & \cellcolor[HTML]{F1E784}0.6074 & \cellcolor[HTML]{FAEA84}0.6065 & \cellcolor[HTML]{FA9E75}0.5310 \\
\hline
PRScs \cite{Ge_2019} & \cellcolor[HTML]{FEDF81}0.1878 & \cellcolor[HTML]{FEE182}0.5717 & \cellcolor[HTML]{E1E383}0.5965 & \cellcolor[HTML]{FDC67C}0.6867 & \cellcolor[HTML]{FFEB84}0.8601 & \cellcolor[HTML]{FFEB84}0.6009 & \cellcolor[HTML]{EEE683}0.6122 & \cellcolor[HTML]{FDD57F}0.5403 \\
\hline
PRScsx \cite{Ruan_2022} & \cellcolor[HTML]{FEEA83}0.2060 & \cellcolor[HTML]{F8E984}0.5817 & \cellcolor[HTML]{FDD47F}0.5675 & \cellcolor[HTML]{FA9573}0.6590 & \cellcolor[HTML]{FEEB84}0.8607 & \cellcolor[HTML]{FCB479}0.5765 & \cellcolor[HTML]{FBAF78}0.5717 & \cellcolor[HTML]{FDD780}0.5407 \\
\hline
PRSice-2 \cite{Choi_2019} & \cellcolor[HTML]{BFD981}0.2669 & \cellcolor[HTML]{B0D580}0.6049 & \cellcolor[HTML]{C5DB81}0.6118 & \cellcolor[HTML]{FEE783}0.7054 & \cellcolor[HTML]{63BE7B}0.9317 & \cellcolor[HTML]{D5DF82}0.6202 & \cellcolor[HTML]{63BE7B}0.6784 & \cellcolor[HTML]{F6E984}0.5494 \\
\hline
RapidoPGS-single \cite{Reales_2021} & \cellcolor[HTML]{FDD27F}0.1654 & \cellcolor[HTML]{FCC47C}0.5498 & \cellcolor[HTML]{FCB87A}0.5515 & \cellcolor[HTML]{FBAF78}0.6736 & \cellcolor[HTML]{FEE983}0.8547 & \cellcolor[HTML]{FDD780}0.5919 & \cellcolor[HTML]{FDD680}0.5928 & \cellcolor[HTML]{FAEA84}0.5470 \\
\hline
SCT \cite{Priv__2019} & \cellcolor[HTML]{EFE784}0.2222 & \cellcolor[HTML]{FCC47C}0.5500 & \cellcolor[HTML]{FDD07E}0.5652 & \cellcolor[HTML]{FDCD7E}0.6907 & \cellcolor[HTML]{FCEB84}0.8616 & \cellcolor[HTML]{D1DE82}0.6220 & \cellcolor[HTML]{D3DF82}0.6251 & \cellcolor[HTML]{F86C6B}0.5225 \\
\hline
SDPR \cite{Zhou_2021} & \cellcolor[HTML]{FEE282}0.1932 & \cellcolor[HTML]{FCB87A}0.5405 & \cellcolor[HTML]{FCC27C}0.5572 & \cellcolor[HTML]{B9D780}0.7280 & \cellcolor[HTML]{FEEA83}0.8590 & \cellcolor[HTML]{FDCE7E}0.5881 & \cellcolor[HTML]{FBA175}0.5640 & \cellcolor[HTML]{FEDB81}0.5414 \\
\hline
smtpred-wMtOLS \cite{Maier_2018} & \cellcolor[HTML]{9BCE7F}0.3014 & \cellcolor[HTML]{C6DB81}0.5978 & \cellcolor[HTML]{FCC27C}0.5572 & \cellcolor[HTML]{A1D07F}0.7350 & \cellcolor[HTML]{FEE983}0.8553 & \cellcolor[HTML]{FBA676}0.5701 & \cellcolor[HTML]{FEDC81}0.5960 & \cellcolor[HTML]{FEE983}0.5437 \\
\hline
smtpred-wMtSBLUP \cite{Maier_2018} & \cellcolor[HTML]{CCDD82}0.2551 & \cellcolor[HTML]{FAA075}0.5221 & \cellcolor[HTML]{FCBD7B}0.5542 & NA& \cellcolor[HTML]{FEE983}0.8549 & NA& \cellcolor[HTML]{FEDC81}0.5959 & \cellcolor[HTML]{FEE783}0.5434 \\
\hline
tlpSum \cite{Pattee_2020} & \cellcolor[HTML]{C2DA81}0.2644 & \cellcolor[HTML]{B2D580}0.6042 & \cellcolor[HTML]{E4E483}0.5948 & \cellcolor[HTML]{FEDC81}0.6992 & \cellcolor[HTML]{FFEB84}0.8603 & \cellcolor[HTML]{F8716C}0.5462 & \cellcolor[HTML]{FDCE7E}0.5886 & \cellcolor[HTML]{FFEB84}0.5444 \\
\hline
VIPRS-Grid \cite{Zabad_2023} & \cellcolor[HTML]{FCC17B}0.1375 & \cellcolor[HTML]{D5DF82}0.5930 & \cellcolor[HTML]{FA9E75}0.5367 & NA& \cellcolor[HTML]{FEE883}0.8542 & NA& \cellcolor[HTML]{CFDE82}0.6270 & \cellcolor[HTML]{FFEB84}0.5440 \\
\hline
VIPRS-Simple \cite{Zabad_2023} & \cellcolor[HTML]{FCC27C}0.1401 & \cellcolor[HTML]{E9E583}0.5867 & \cellcolor[HTML]{FDC77D}0.5601 & NA& \cellcolor[HTML]{FEE883}0.8539 & NA& \cellcolor[HTML]{DAE182}0.6217 & \cellcolor[HTML]{FEE683}0.5432 \\
\hline
XPBLUP \cite{Zhang_2023} & \cellcolor[HTML]{F86F6C}0.0025 & \cellcolor[HTML]{FEE081}0.5710 & \cellcolor[HTML]{FCB77A}0.5509 & \cellcolor[HTML]{E2E383}0.7158 & \cellcolor[HTML]{8ECB7E}0.9122 & \cellcolor[HTML]{DAE182}0.6178 & \cellcolor[HTML]{91CC7E}0.6566 & \cellcolor[HTML]{63BE7B}0.6291 \\
\hline
\end{tabular}%
}
\caption{Best predictive performance achieved by each PRS tool across the eight phenotypes, reported as test-set performance for the full model. Values for Height are reported as $R^2$, whereas values for all binary phenotypes are reported as AUC. Each entry corresponds to the selected full-model configuration for that tool and phenotype, and \texttt{NA} indicates that no valid result was obtained for that phenotype. Cell colours represent relative performance within each phenotype column.}
\label{TableBestFullModelSorted}
\end{table*}

Predictive performance varied substantially across both tools and phenotypes, indicating that no single PRS method consistently achieved the strongest results in all settings. LDAK-GWAS performed best for Height ($R^2$~=~0.3531), LDpred-2-Grid for Asthma (AUC~=~0.6295), LDAK-GWAS for Depression (AUC~=~0.6637), BOLT-LMM for Gastro-Reflux (AUC~=~0.7528), PRSice-2 for High Cholesterol (AUC~=~0.9317), LDpred-inf for Hypothyroidism (AUC~=~0.6710), PRSice-2 for Irritable Bowel Syndrome (IBS) (AUC~=~0.6784), and XPBLUP for Migraine (AUC~=~0.6291). These results are consistent with previous observations that PRS tool performance is phenotype-dependent and influenced by trait architecture and evaluation design~\cite{Ma2021}.

Although several methods performed strongly across multiple phenotypes, the relative ranking of tools changed markedly by phenotype. This phenotype dependence was also reflected in the distribution of top-performing tools across phenotypes, with no single method dominating all analyses. Taken together, these findings support the interpretation of the benchmark as a comparative multi-phenotype evaluation framework rather than a definitive ranking of PRS tools.

\subsection*{PRS improve prediction beyond the null model in selected phenotypes}

To assess whether PRS contributed predictive value beyond baseline covariates, we compared the full model (PRS + covariates + principal components) with a phenotype-specific null model that included only covariates and principal components. For each tool and phenotype, hyperparameter configurations were selected using the predefined $\delta$-constrained rule. The resulting tool-specific improvement values are summarized in Table~\ref{TablePRSImprovement}. Because the binary phenotype models included a rich covariate structure, these comparisons should be interpreted as the incremental contribution of PRS within an information-rich prediction setting rather than as a pure test of genetic prediction in isolation.

\begin{table*}[!ht]
\centering
\resizebox{\textwidth}{!}{%
\begin{tabular}{|l|l|l|l|l|l|l|l|l|}
\hline
\textbf{Tool} & \textbf{Height} & \textbf{Asthma} & \textbf{Depression} & \textbf{Gastro-Reflux} & \textbf{High Cholesterol} & \textbf{Hypothyroidism} & \textbf{IBS} & \textbf{Migraine} \\
\hline
\textbf{AnnoPred} & \cellcolor[HTML]{ACD380}0.1833 & \cellcolor[HTML]{77C47D}0.0570 & \cellcolor[HTML]{D3DF82}0.0522 & NA& \cellcolor[HTML]{FEE883}0.0007 & NA& \cellcolor[HTML]{FCB579}-0.0176 & \cellcolor[HTML]{FBAC78}-0.0093 \\
\hline
\textbf{BOLT-LMM} & \cellcolor[HTML]{FA8F72}-0.1586 & \cellcolor[HTML]{FBA476}-0.0005 & \cellcolor[HTML]{FCB379}-0.0001 & \cellcolor[HTML]{68C07C}0.0372 & \cellcolor[HTML]{D0DE82}0.0277 & \cellcolor[HTML]{63BE7B}0.0376 & \cellcolor[HTML]{ECE683}0.0181 & \cellcolor[HTML]{8ACA7E}0.0464 \\
\hline
\textbf{C+T} & \cellcolor[HTML]{90CB7E}0.2176 & \cellcolor[HTML]{DFE283}0.0369 & \cellcolor[HTML]{F3E884}0.0353 & \cellcolor[HTML]{FFEB84}-0.0054 & \cellcolor[HTML]{FEEB84}0.0067 & \cellcolor[HTML]{FEE182}-0.0287 & \cellcolor[HTML]{FDD37F}-0.0033 & \cellcolor[HTML]{FEE783}-0.0003 \\
\hline
\textbf{CTPR} & \cellcolor[HTML]{FEDB81}0.0393 & \cellcolor[HTML]{F97D6E}-0.0178 & \cellcolor[HTML]{D0DE82}0.0541 & \cellcolor[HTML]{FEE382}-0.0077 & \cellcolor[HTML]{F6E984}0.0105 & \cellcolor[HTML]{B1D580}0.0066 & \cellcolor[HTML]{FA9C74}-0.0291 & \cellcolor[HTML]{FEE683}-0.0005 \\
\hline
\textbf{CTSLEB} & \cellcolor[HTML]{FDD27F}0.0143 & \cellcolor[HTML]{FBA175}-0.0019 & \cellcolor[HTML]{EEE683}0.0380 & \cellcolor[HTML]{FA9473}-0.0290 & \cellcolor[HTML]{EFE784}0.0135 & \cellcolor[HTML]{DDE182}-0.0106 & \cellcolor[HTML]{C9DC81}0.0368 & \cellcolor[HTML]{F8706C}-0.0185 \\
\hline
\textbf{DBSLMM} & \cellcolor[HTML]{FDCC7E}0.0000 & \cellcolor[HTML]{A1D07F}0.0489 & \cellcolor[HTML]{A4D17F}0.0774 & \cellcolor[HTML]{FFEB84}-0.0056 & \cellcolor[HTML]{FEEB84}0.0065 & \cellcolor[HTML]{FDD07E}-0.0366 & \cellcolor[HTML]{FDD780}-0.0013 & \cellcolor[HTML]{FFEB84}0.0004 \\
\hline
\textbf{EBPRS} & \cellcolor[HTML]{EFE784}0.0989 & \cellcolor[HTML]{DAE182}0.0379 & \cellcolor[HTML]{D7E082}0.0503 & \cellcolor[HTML]{F87B6E}-0.0356 & \cellcolor[HTML]{FEE883}0.0007 & \cellcolor[HTML]{F8696B}-0.0835 & \cellcolor[HTML]{FDCE7E}-0.0055 & \cellcolor[HTML]{E1E383}0.0120 \\
\hline
\textbf{GCTA} & \cellcolor[HTML]{C6DB81}0.1510 & \cellcolor[HTML]{FDC67D}0.0146 & \cellcolor[HTML]{B0D580}0.0709 & \cellcolor[HTML]{FEEB84}-0.0051 & \cellcolor[HTML]{F8696B}-0.2731 & \cellcolor[HTML]{E3E383}-0.0131 & \cellcolor[HTML]{DDE283}0.0261 & \cellcolor[HTML]{FA9874}-0.0123 \\
\hline
\textbf{GCTB-SBayesR} & \cellcolor[HTML]{FDEB84}0.0819 & \cellcolor[HTML]{FDD680}0.0218 & \cellcolor[HTML]{FCC37C}0.0083 & \cellcolor[HTML]{F3E884}-0.0021 & \cellcolor[HTML]{FFEB84}0.0062 & \cellcolor[HTML]{FEE282}-0.0284 & \cellcolor[HTML]{FBA476}-0.0255 & \cellcolor[HTML]{FEE482}-0.0008 \\
\hline
\textbf{GCTB-SBayesRC} & \cellcolor[HTML]{FEDD81}0.0438 & \cellcolor[HTML]{FBB179}0.0054 & \cellcolor[HTML]{FEDD81}0.0219 & \cellcolor[HTML]{FEDA80}-0.0100 & \cellcolor[HTML]{EEE784}0.0138 & \cellcolor[HTML]{FBB078}-0.0512 & \cellcolor[HTML]{F98570}-0.0401 & \cellcolor[HTML]{FEEA83}0.0001 \\
\hline
\textbf{GEMMA-BSLMM} & \cellcolor[HTML]{FDC97D}-0.0096 & \cellcolor[HTML]{FEE182}0.0266 & \cellcolor[HTML]{F8E984}0.0330 & \cellcolor[HTML]{FEDA80}-0.0100 & \cellcolor[HTML]{FEEB84}0.0069 & \cellcolor[HTML]{FCC57C}-0.0415 & \cellcolor[HTML]{FEDD81}0.0015 & \cellcolor[HTML]{F2E884}0.0055 \\
\hline
\textbf{GEMMA-LM} & \cellcolor[HTML]{F8696B}-0.2601 & \cellcolor[HTML]{E9E583}0.0350 & \cellcolor[HTML]{8BCA7E}0.0904 & \cellcolor[HTML]{FEE482}-0.0074 & \cellcolor[HTML]{A8D27F}0.0457 & \cellcolor[HTML]{FBAF78}-0.0517 & \cellcolor[HTML]{FA9F75}-0.0280 & \cellcolor[HTML]{63BE7B}0.0615 \\
\hline
\textbf{GEMMA-LMM} & \cellcolor[HTML]{F98570}-0.1865 & \cellcolor[HTML]{FBB178}0.0053 & \cellcolor[HTML]{FCC57C}0.0092 & \cellcolor[HTML]{D4DF82}0.0067 & \cellcolor[HTML]{ECE683}0.0150 & \cellcolor[HTML]{FBAA77}-0.0539 & \cellcolor[HTML]{E3E383}0.0231 & \cellcolor[HTML]{7CC67D}0.0517 \\
\hline
\textbf{HAIL} & \cellcolor[HTML]{FDD57F}0.0221 & \cellcolor[HTML]{F86F6C}-0.0241 & \cellcolor[HTML]{AED480}0.0721 & \cellcolor[HTML]{FBB178}-0.0212 & \cellcolor[HTML]{BFD981}0.0352 & \cellcolor[HTML]{F4E884}-0.0200 & \cellcolor[HTML]{FEE282}0.0040 & \cellcolor[HTML]{FCC47C}-0.0057 \\
\hline
\textbf{JAMPred} & \cellcolor[HTML]{FCBA7A}-0.0484 & \cellcolor[HTML]{FEE582}0.0281 & \cellcolor[HTML]{FA9072}-0.0185 & NA& \cellcolor[HTML]{FEE883}0.0015 & NA& \cellcolor[HTML]{FEE883}0.0068 & \cellcolor[HTML]{E2E383}0.0119 \\
\hline
\textbf{LDAK-GenotypeData} & \cellcolor[HTML]{FA9C74}-0.1257 & \cellcolor[HTML]{F98D72}-0.0105 & \cellcolor[HTML]{F0E784}0.0372 & \cellcolor[HTML]{F9EA84}-0.0039 & \cellcolor[HTML]{FFEB84}0.0060 & \cellcolor[HTML]{FEDE81}-0.0302 & \cellcolor[HTML]{C6DB81}0.0382 & \cellcolor[HTML]{FEE582}-0.0006 \\
\hline
\textbf{LDAK-GWAS} & \cellcolor[HTML]{63BE7B}0.2739 & \cellcolor[HTML]{FA8F72}-0.0096 & \cellcolor[HTML]{FDD780}0.0187 & \cellcolor[HTML]{FDD17F}-0.0124 & \cellcolor[HTML]{FEEB84}0.0065 & \cellcolor[HTML]{FEE482}-0.0273 & \cellcolor[HTML]{FED880}-0.0007 & \cellcolor[HTML]{FFEB84}0.0002 \\
\hline
\textbf{Lassosum} & \cellcolor[HTML]{87C97E}0.2290 & \cellcolor[HTML]{9FD07F}0.0493 & \cellcolor[HTML]{C8DC81}0.0582 & \cellcolor[HTML]{FBA175}-0.0254 & \cellcolor[HTML]{FAEA84}0.0085 & \cellcolor[HTML]{E6E483}-0.0145 & \cellcolor[HTML]{FDD780}-0.0013 & \cellcolor[HTML]{FEE983}-0.0001 \\
\hline
\textbf{LDpred-2-Auto} & \cellcolor[HTML]{B4D680}0.1727 & \cellcolor[HTML]{FCC57C}0.0141 & \cellcolor[HTML]{FEE081}0.0233 & NA& \cellcolor[HTML]{FEEA83}0.0059 & \cellcolor[HTML]{E6E483}-0.0142 & \cellcolor[HTML]{FDC87D}-0.0083 & NA\\
\hline
\textbf{LDpred-2-Grid} & \cellcolor[HTML]{FEE983}0.0751 & \cellcolor[HTML]{8ACA7E}0.0532 & \cellcolor[HTML]{FDD67F}0.0179 & \cellcolor[HTML]{FCB379}-0.0206 & \cellcolor[HTML]{FEEA83}0.0057 & \cellcolor[HTML]{FDD07E}-0.0366 & \cellcolor[HTML]{F98871}-0.0388 & NA\\
\hline
\textbf{LDpred-2-Inf} & \cellcolor[HTML]{D1DE82}0.1369 & \cellcolor[HTML]{9DCF7F}0.0497 & \cellcolor[HTML]{FEE482}0.0255 & \cellcolor[HTML]{FEDF81}-0.0087 & \cellcolor[HTML]{FEEA83}0.0050 & \cellcolor[HTML]{FCB579}-0.0486 & \cellcolor[HTML]{D8E082}0.0288 & NA\\
\hline
\textbf{LDpred-2-Lassosum2} & \cellcolor[HTML]{C6DB81}0.1502 & \cellcolor[HTML]{7DC67D}0.0558 & \cellcolor[HTML]{FAEA84}0.0318 & \cellcolor[HTML]{D4DF82}0.0068 & \cellcolor[HTML]{FEEA83}0.0056 & \cellcolor[HTML]{FCC27C}-0.0431 & \cellcolor[HTML]{EBE583}0.0191 & \cellcolor[HTML]{FEEB84}0.0009 \\
\hline
\textbf{LDpred-fast} & \cellcolor[HTML]{86C87D}0.2313 & \cellcolor[HTML]{AAD380}0.0472 & \cellcolor[HTML]{FBAE78}-0.0026 & \cellcolor[HTML]{FBAC77}-0.0225 & NA& \cellcolor[HTML]{FDD680}-0.0339 & \cellcolor[HTML]{FAEA84}0.0108 & \cellcolor[HTML]{FEEB84}0.0009 \\
\hline
\textbf{LDpred-funct} & \cellcolor[HTML]{D3DF82}0.1347 & \cellcolor[HTML]{FEE783}0.0289 & \cellcolor[HTML]{FCB679}0.0015 & NA& \cellcolor[HTML]{FEE883}0.0013 & NA& \cellcolor[HTML]{FFEB84}0.0080 & \cellcolor[HTML]{F5E884}0.0043 \\
\hline
\textbf{LDpred-gibbs} & \cellcolor[HTML]{C4DA81}0.1531 & \cellcolor[HTML]{E4E483}0.0359 & \cellcolor[HTML]{F98D72}-0.0198 & \cellcolor[HTML]{C1D981}0.0121 & \cellcolor[HTML]{FEE883}0.0011 & \cellcolor[HTML]{FEE783}-0.0261 & \cellcolor[HTML]{F8766D}-0.0473 & \cellcolor[HTML]{FFEB84}0.0003 \\
\hline
\textbf{LDpred-inf} & \cellcolor[HTML]{95CD7E}0.2124 & \cellcolor[HTML]{F8696B}-0.0268 & \cellcolor[HTML]{F98770}-0.0229 & \cellcolor[HTML]{E4E483}0.0021 & \cellcolor[HTML]{FEE883}0.0011 & \cellcolor[HTML]{ACD380}0.0089 & \cellcolor[HTML]{F7E984}0.0125 & \cellcolor[HTML]{FEE983}0.0000 \\
\hline
\textbf{LDpred-p+t} & \cellcolor[HTML]{81C77D}0.2366 & \cellcolor[HTML]{63BE7B}0.0607 & \cellcolor[HTML]{B9D780}0.0661 & \cellcolor[HTML]{63BE7B}0.0385 & \cellcolor[HTML]{FEE883}0.0014 & \cellcolor[HTML]{E8E583}-0.0152 & \cellcolor[HTML]{F9EA84}0.0114 & NA\\
\hline
\textbf{MTG2} & \cellcolor[HTML]{F98670}-0.1829 & \cellcolor[HTML]{FA9573}-0.0072 & \cellcolor[HTML]{E8E583}0.0411 & NA& \cellcolor[HTML]{81C77D}0.0634 & \cellcolor[HTML]{C9DC81}-0.0028 & \cellcolor[HTML]{64BF7C}0.0901 & \cellcolor[HTML]{73C37C}0.0555 \\
\hline
\textbf{NPS} & \cellcolor[HTML]{FDD680}0.0264 & \cellcolor[HTML]{FEE683}0.0285 & \cellcolor[HTML]{FBAD78}-0.0032 & NA& NA& NA& \cellcolor[HTML]{FBEA84}0.0105 & \cellcolor[HTML]{FA9673}-0.0127 \\
\hline
\textbf{PANPRS} & \cellcolor[HTML]{FEE582}0.0640 & \cellcolor[HTML]{CCDD82}0.0405 & \cellcolor[HTML]{FCBD7B}0.0051 & \cellcolor[HTML]{E1E383}0.0029 & \cellcolor[HTML]{FEE883}0.0014 & \cellcolor[HTML]{A3D17F}0.0125 & \cellcolor[HTML]{FBA276}-0.0264 & \cellcolor[HTML]{DEE283}0.0135 \\
\hline
\textbf{PleioPred} & \cellcolor[HTML]{86C87D}0.2311 & \cellcolor[HTML]{EAE583}0.0347 & NA& NA& NA& NA& NA& NA\\
\hline
\textbf{PLINK} & \cellcolor[HTML]{91CC7E}0.2166 & \cellcolor[HTML]{B8D780}0.0444 & \cellcolor[HTML]{FBA877}-0.0059 & \cellcolor[HTML]{ECE683}0.0000 & \cellcolor[HTML]{FEE883}0.0006 & \cellcolor[HTML]{FEEB84}-0.0241 & \cellcolor[HTML]{FBEA84}0.0105 & \cellcolor[HTML]{F7E984}0.0034 \\
\hline
\textbf{PolyPred} & \cellcolor[HTML]{FDCB7D}-0.0028 & \cellcolor[HTML]{B2D580}0.0455 & \cellcolor[HTML]{63BE7B}0.1116 & NA& \cellcolor[HTML]{F9EA84}0.0091 & NA& \cellcolor[HTML]{A0D07F}0.0583 & \cellcolor[HTML]{92CC7E}0.0432 \\
\hline
\textbf{PRSbils} & \cellcolor[HTML]{FEDB80}0.0374 & \cellcolor[HTML]{FCBF7B}0.0115 & \cellcolor[HTML]{FA9673}-0.0153 & \cellcolor[HTML]{FEDD81}-0.0092 & \cellcolor[HTML]{FEEA83}0.0053 & \cellcolor[HTML]{FEDD81}-0.0306 & \cellcolor[HTML]{FCEA84}0.0101 & \cellcolor[HTML]{F98370}-0.0155 \\
\hline
\textbf{PRScs} & \cellcolor[HTML]{F7E984}0.0890 & \cellcolor[HTML]{CEDD82}0.0402 & \cellcolor[HTML]{EDE683}0.0387 & \cellcolor[HTML]{F8756D}-0.0372 & \cellcolor[HTML]{FFEB84}0.0064 & \cellcolor[HTML]{F3E884}-0.0195 & \cellcolor[HTML]{E8E583}0.0206 & \cellcolor[HTML]{FEE683}-0.0005 \\
\hline
\textbf{PRScsx} & \cellcolor[HTML]{EFE784}0.0994 & \cellcolor[HTML]{FA9272}-0.0086 & \cellcolor[HTML]{FFEB84}0.0289 & \cellcolor[HTML]{F8696B}-0.0406 & \cellcolor[HTML]{FEEB84}0.0067 & \cellcolor[HTML]{F8E984}-0.0217 & \cellcolor[HTML]{FEE482}0.0051 & \cellcolor[HTML]{FEEA83}0.0001 \\
\hline
\textbf{PRSice-2} & \cellcolor[HTML]{B2D580}0.1758 & \cellcolor[HTML]{ABD380}0.0469 & \cellcolor[HTML]{B7D780}0.0673 & \cellcolor[HTML]{F9EA84}-0.0037 & \cellcolor[HTML]{63BE7B}0.0770 & \cellcolor[HTML]{F1E784}-0.0188 & \cellcolor[HTML]{63BE7B}0.0905 & \cellcolor[HTML]{EFE784}0.0067 \\
\hline
\textbf{RapidoPGS-single} & \cellcolor[HTML]{FEDE81}0.0450 & \cellcolor[HTML]{7EC67D}0.0555 & \cellcolor[HTML]{FBA676}-0.0070 & \cellcolor[HTML]{C7DB81}0.0104 & \cellcolor[HTML]{FEE883}0.0013 & \cellcolor[HTML]{FCC17B}-0.0435 & \cellcolor[HTML]{D7E082}0.0292 & \cellcolor[HTML]{F3E884}0.0051 \\
\hline
\textbf{SCT} & \cellcolor[HTML]{FED980}0.0326 & \cellcolor[HTML]{FCC37C}0.0130 & \cellcolor[HTML]{F8696B}-0.0390 & \cellcolor[HTML]{FDCB7E}-0.0140 & \cellcolor[HTML]{FAEA84}0.0084 & \cellcolor[HTML]{D5DF82}-0.0075 & \cellcolor[HTML]{D5DF82}0.0306 & \cellcolor[HTML]{F8696B}-0.0196 \\
\hline
\textbf{SDPR} & \cellcolor[HTML]{FEDF81}0.0474 & \cellcolor[HTML]{FA9D75}-0.0034 & \cellcolor[HTML]{FCB87A}0.0026 & \cellcolor[HTML]{E0E283}0.0034 & \cellcolor[HTML]{FEE983}0.0035 & \cellcolor[HTML]{7EC67D}0.0270 & \cellcolor[HTML]{FA9974}-0.0309 & \cellcolor[HTML]{FEEA83}0.0001 \\
\hline
\textbf{smtpred-wMtOLS} & \cellcolor[HTML]{95CD7E}0.2113 & \cellcolor[HTML]{82C77D}0.0549 & \cellcolor[HTML]{FBAD78}-0.0033 & \cellcolor[HTML]{C4DA81}0.0111 & \cellcolor[HTML]{FEE883}0.0013 & \cellcolor[HTML]{BAD780}0.0031 & \cellcolor[HTML]{FCBE7B}-0.0132 & \cellcolor[HTML]{FFEB84}0.0002 \\
\hline
\textbf{smtpred-wMtSBLUP} & \cellcolor[HTML]{C3DA81}0.1544 & \cellcolor[HTML]{FEE983}0.0301 & \cellcolor[HTML]{EDE683}0.0387 & NA& \cellcolor[HTML]{FEE883}0.0010 & NA& \cellcolor[HTML]{F8696B}-0.0537 & \cellcolor[HTML]{FEE683}-0.0005 \\
\hline
\textbf{tlpSum} & \cellcolor[HTML]{A5D17F}0.1919 & \cellcolor[HTML]{A8D27F}0.0476 & \cellcolor[HTML]{EEE784}0.0379 & \cellcolor[HTML]{EBE583}0.0003 & \cellcolor[HTML]{FEEB84}0.0069 & \cellcolor[HTML]{FEEA83}-0.0248 & \cellcolor[HTML]{FDD27F}-0.0038 & \cellcolor[HTML]{FEE983}0.0000 \\
\hline
\textbf{VIPRS-Grid} & \cellcolor[HTML]{FDCC7E}-0.0001 & \cellcolor[HTML]{FDCF7E}0.0183 & \cellcolor[HTML]{F3E884}0.0353 & NA& \cellcolor[HTML]{FEE883}0.0008 & NA& \cellcolor[HTML]{DDE283}0.0260 & \cellcolor[HTML]{FFEB84}0.0003 \\
\hline
\textbf{VIPRS-Simple} & \cellcolor[HTML]{FDCD7E}0.0020 & \cellcolor[HTML]{FDEB84}0.0312 & \cellcolor[HTML]{FDC67D}0.0099 & NA& \cellcolor[HTML]{FEE883}0.0007 & NA& \cellcolor[HTML]{D3DF82}0.0314 & \cellcolor[HTML]{FEEA83}0.0001 \\
\hline
\textbf{XPBLUP} & \cellcolor[HTML]{F98871}-0.1770 & \cellcolor[HTML]{88C97E}0.0536 & \cellcolor[HTML]{FFEB84}0.0288 & \cellcolor[HTML]{FBA175}-0.0253 & \cellcolor[HTML]{93CC7E}0.0556 & \cellcolor[HTML]{FFEB84}-0.0245 & \cellcolor[HTML]{EEE683}0.0175 & \cellcolor[HTML]{77C47D}0.0538 \\
\hline
\end{tabular}%
}
\caption{Improvement in predictive performance of the full model relative to the phenotype-specific null model across PRS tools and phenotypes. For Height, values represent the difference in explained variance ($\Delta R^2$), whereas for all binary phenotypes, values represent the difference in test AUC ($\Delta$AUC). Positive values indicate that the inclusion of the PRS improved prediction beyond that of covariates and principal components alone, whereas negative values indicate reduced performance relative to the null model. Missing values (\texttt{NA}) indicate that no valid result was available for that tool--phenotype combination. Cell colours represent relative improvement patterns within each phenotype column.}
\label{TablePRSImprovement}
\end{table*}

Positive values in Table~\ref{TablePRSImprovement} indicate improved predictive performance after inclusion of the PRS component, whereas negative values indicate that adding the PRS did not improve performance relative to the null model. PRS-related improvement varied substantially across phenotypes and tools. The largest improvement was observed for Height, where LDAK-GWAS increased explained variance from 0.128 in the null model to 0.353 in the full model ($\Delta = 0.225$). Among the binary phenotypes, the largest gain was observed for Depression, where LDAK-GWAS increased test AUC from 0.524 to 0.664 ($\Delta = 0.140$), followed by Asthma, where LDpred-2-Grid increased test AUC from 0.504 to 0.630 ($\Delta = 0.126$). For IBS, PRSice-2 improved test AUC from 0.589 to 0.678 ($\Delta = 0.089$), and for Migraine, XPBLUP increased test AUC from 0.540 to 0.629 ($\Delta = 0.089$).

To evaluate whether these improvements were statistically consistent across folds, we compared null and full model performance using paired Wilcoxon signed-rank tests across all 340 phenotype--tool comparisons. No comparison reached nominal significance at $p < 0.05$, and none remained significant after false discovery rate correction ($q < 0.05$). For the phenotype-level best-performing full models, the observed p-value was 0.0625 for Height, Asthma, Depression, High Cholesterol, IBS, and Migraine, and 0.3125 for Gastro-Reflux and Hypothyroidism (Table~\ref{TablePRSvsNullSummary}). The corresponding FDR q-values were 0.2043 and 0.5183, respectively. These results should be interpreted with caution: with only five cross-validation folds, the minimum attainable two-sided Wilcoxon p-value is 0.0625, implying that formal significance at $p < 0.05$ is mathematically unattainable under this design, regardless of the true effect size. The p-values reported in Table~\ref{TablePRSvsNullSummary} therefore represent the most extreme evidence the fold count permits rather than weak effects. The direction of improvement was consistent across folds for six of eight phenotypes, and the effect sizes ($\Delta$) were practically meaningful for Height ($\Delta = 0.225$), Depression ($\Delta = 0.140$), and Asthma ($\Delta = 0.126$). We therefore interpret these results as directionally supportive of PRS utility, while acknowledging that definitive statistical confirmation would require a larger number of folds or an independent replication cohort.

\begin{table}[!ht]
\centering
\begin{tabular}{|l|l|l|l|l|l|l|}
\hline
\textbf{Phenotype} & \textbf{Null} & \textbf{Best Tool} & \textbf{Full} & \textbf{$\Delta$} & \textbf{p-value} & \textbf{FDR q-value} \\
\hline
Height & 0.128 & LDAK-GWAS & 0.353 & 0.225 & 0.0625 & 0.2043 \\
\hline
Asthma & 0.504 & LDpred-2-Grid & 0.630 & 0.126 & 0.0625 & 0.2043 \\
\hline
Depression & 0.524 & LDAK-GWAS & 0.664 & 0.140 & 0.0625 & 0.2043 \\
\hline
Gastro-Reflux & 0.705 & GEMMA-BSLMM & 0.743 & 0.038 & 0.3125 & 0.5183 \\
\hline
High Cholesterol & 0.853 & PRSice-2 & 0.932 & 0.079 & 0.0625 & 0.2043 \\
\hline
Hypothyroidism & 0.629 & LDpred-inf & 0.671 & 0.042 & 0.3125 & 0.5183 \\
\hline
IBS & 0.589 & PRSice-2 & 0.678 & 0.089 & 0.0625 & 0.2043 \\
\hline
Migraine & 0.540 & XPBLUP & 0.629 & 0.089 & 0.0625 & 0.2043 \\
\hline
\end{tabular}
\caption{Phenotype-level comparison between the null model and the best-performing full model. For each phenotype, the best-performing tool was defined as the method with the largest mean test improvement relative to the null model. $\Delta$ represents the difference between the mean test performance of the full and null models. P-values were obtained using paired Wilcoxon signed-rank tests across the five cross-validation folds, and FDR q-values were calculated across all phenotype--tool comparisons. Note that with five folds, the minimum attainable two-sided Wilcoxon p-value is 0.0625, so reported p-values reflect the power ceiling of the cross-validation design rather than the absence of effect.}
\label{TablePRSvsNullSummary}
\end{table}

These results show that PRS can improve prediction beyond covariate-only models for several phenotypes, particularly Height, Depression, Asthma, High Cholesterol, IBS, and Migraine, although the magnitude of this gain was phenotype-specific. In contrast, the improvements observed for Gastro-Reflux and Hypothyroidism were smaller and less consistent across folds. Complete null, pure-PRS, and full-model comparisons are provided in Supplementary Table 4 (Train Test Null-Pure-Full).

\subsection*{Global statistical comparison identifies consistently strong performers}

To assess whether predictive performance differed systematically across PRS tools, we performed a global repeated-measures comparison using fold-level test performance from the full model. For each phenotype and cross-validation fold, the best-performing configuration for each tool was selected using the predefined $\delta$-constrained rule, and the tools were ranked by their test-set performance. This produced 40 phenotype--fold datasets (8 phenotypes $\times$ 5 folds), enabling comparison of PRS tools across a common benchmarking framework.

The Friedman test showed a statistically significant difference in predictive performance among PRS tools across these phenotype--fold combinations ($\chi^2 = 102.29$, $p = 2.57 \times 10^{-11}$), indicating that tool rankings were not equivalent across the evaluated settings. Average-rank analysis identified a subset of methods that performed consistently well across phenotypes. Global statistical comparison identified \textit{LDpred-2-Lassosum2} as the most consistently strong performer (average rank = 9.54), followed by \textit{PRSice-2} (9.80) and \textit{LDAK-GWAS} (10.22). Other methods that ranked among the strongest overall performers included \textit{PANPRS}, \textit{GEMMA-LMM}, \textit{Lassosum}, \textit{GEMMA-LM}, \textit{XPBLUP}, \textit{C+T}, and \textit{LDpred-inf}. Post hoc Nemenyi comparisons identified 34 pairwise differences that exceeded the critical-difference threshold, and 93 pairwise comparisons remained significant after Benjamini--Hochberg false discovery rate correction. LDpred-2-Lassosum2 showed the strongest overall performance under this framework, with 10 statistically significant pairwise wins under the critical-difference criterion and 15 wins after FDR correction. Nevertheless, no single method dominated across all phenotypes, indicating that a strong global ranking does not imply universal superiority across all traits.

Overall, these results demonstrate that PRS tool choice has a substantial effect on predictive performance and that several methods consistently outperform others across diverse phenotypes. At the same time, the absence of a universally dominant method reinforces the importance of benchmarking multiple tools rather than relying on a single default approach when constructing polygenic risk scores. The robustness of these rankings to the choice of hyperparameter selection rule is evaluated in the following subsection.


\subsection*{Sensitivity analysis reveals overfitting-prone tools and validates the stability-constrained selection rule} 

The Friedman test confirmed statistically significant differences in tool rankings under all four selection rules, demonstrating that the finding of significant performance heterogeneity across PRS tools is not an artefact of the selection strategy. The stricter and more lenient $\delta$ variants produced rankings nearly identical to the main analysis (Spearman $\rho = 0.972$ and $\rho = 0.976$ respectively, both $p < 10^{-28}$; top-10 overlap = 10/10 and 9/10 tools respectively), confirming that conclusions are insensitive to the specific $\delta$ threshold chosen. 

Comparison with the training-only rule yielded a lower Spearman correlation ($\rho = 0.34$, $p = 0.021$) and a top-10 overlap of 4/10 tools. Rather than indicating a weakness of the main selection rule, this divergence is itself informative about the overfitting behaviour of specific PRS methods. Tools that ranked substantially worse under the training-only condition --- including GEMMA-LMM (average rank 16.3 under the main rule versus 25.2 under train-only), GEMMA-LM (17.0 versus 25.3), and MTG2 (17.8 versus 28.5) --- rely on full genotype-based LD modelling or linear mixed-model estimation, where configurations optimised for training performance do not generalise reliably to held-out test sets. By contrast, tools that maintained strong rankings under both rules --- including PRSice-2 (average rank 14.3 under the main rule and 11.9 under train-only, remaining the top-ranked tool in both cases), Lassosum, and LDpred-p+t --- are summary-statistics-based or clumping methods with inherent regularisation, whose selected configurations generalise robustly regardless of whether the stability criterion is applied. These results demonstrate two complementary properties of the $\delta$-constrained rule: it is insensitive to the exact threshold chosen, and it actively protects against overfitting in genotype-dependent methods where the stability constraint is most consequential. PRSice-2, Lassosum, and LDpred-p+t were the only three tools that ranked consistently among the top 10 across all four selection conditions, representing the most robust performers in the benchmark under any selection strategy.

\subsection*{Performance and operational complexity define four distinct tool profiles} 

\begin{figure}[!ht]
\centering 
\includegraphics[width=\textwidth]{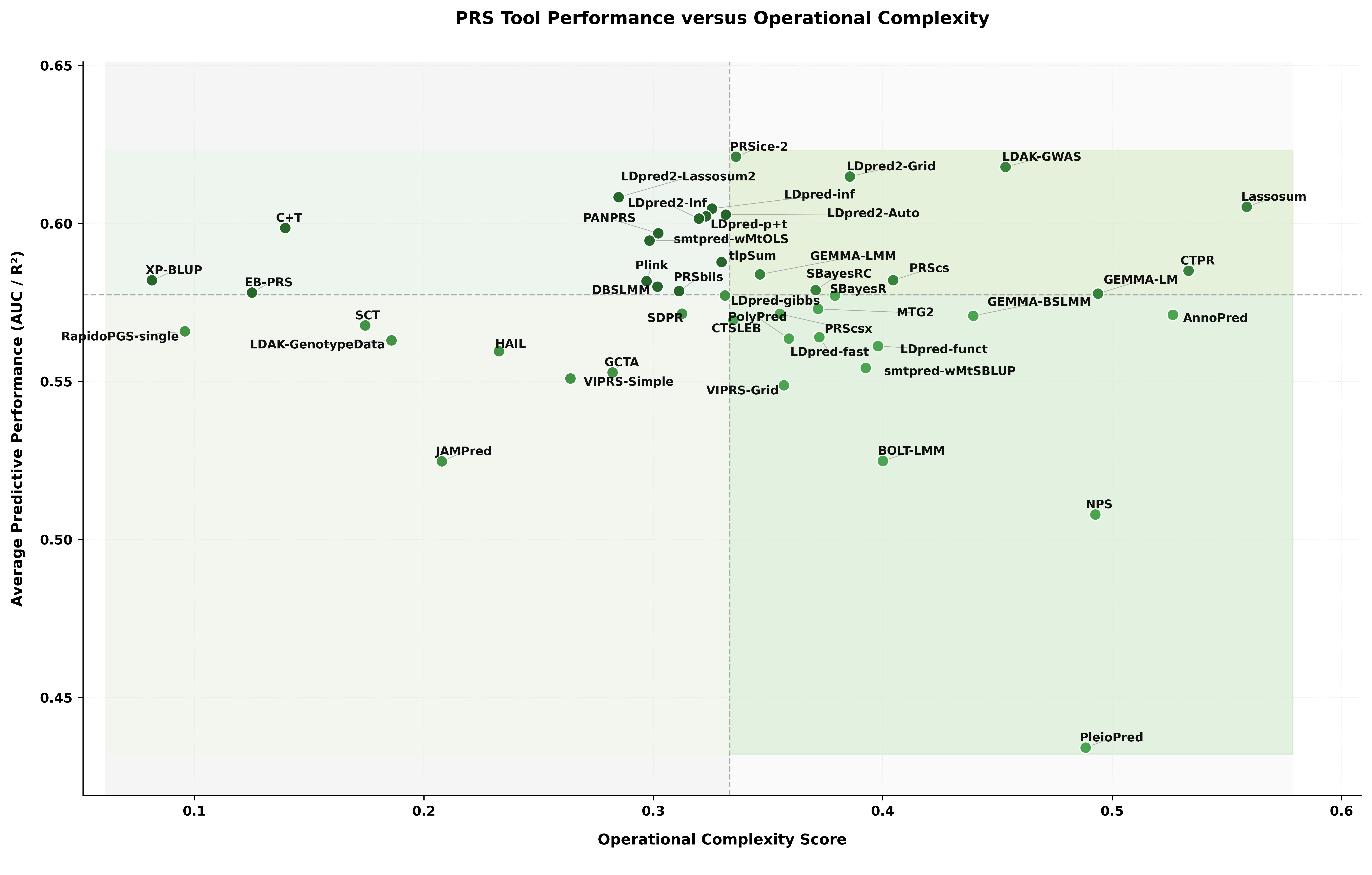} \caption{\textbf{Predictive performance versus operational complexity across 46 PRS tools.} Each point represents one PRS tool. The x-axis shows a composite operational complexity score computed as a weighted sum of five normalised components: data input requirements ($w = 0.20$), LD modelling burden ($w = 0.15$), log-normalised mean runtime ($w = 0.25$), normalised mean memory consumption ($w = 0.15$), and phenotype-level failure rate ($w = 0.25$), where failure rate is defined as the proportion of 40 phenotype--fold combinations for which no valid result was produced. The y-axis shows average predictive performance across all phenotypes for which the tool produced a valid result, expressed as AUC for binary phenotypes and $R^2$ for the continuous Height phenotype and averaged across all evaluated traits. Dashed lines denote the median complexity score (0.33) and median performance (0.578), dividing the space into four quadrants. Quadrant background shading reflects the mean performance of tools within each quadrant, with deeper shading indicating higher average performance.} \label{fig:performance_burden} 
\end{figure}

To provide an integrated view of PRS tool utility beyond predictive accuracy alone, we constructed a composite operational complexity score for each tool and plotted it against average predictive performance (Figure~\ref{fig:performance_burden}). The complexity score integrates five components: data input requirements (weight = 0.20), scored across three input types --- genotype data, GWAS summary statistics, and reference panels --- as required (1), optional (0.5), or not needed (0), and normalised to 0--1; LD modelling burden (0.15), scored as 0 for no LD model, 0.5 for block or banded matrices, and 1.0 for full genomic relationship matrices; log-normalised mean runtime (0.25); normalised mean memory consumption (0.15); and phenotype-level failure rate (0.25). Failure rate was defined as the proportion of the 40 phenotype--fold combinations for which a tool did not produce a valid result, and was assigned the highest weight alongside runtime because it directly limits deployability in multi-phenotype analyses. The five components were each normalised to a 0--1 scale before combination, and the weights were chosen to reflect the practical burden each dimension imposes in a real HPC benchmarking environment. 

The median complexity score (0.33) and median performance (0.578) divide tools into four quadrants. Tools in the upper-left quadrant combine above-median predictive performance with below-median operational complexity, and include C+T, XP-BLUP, LDpred2-Lassosum2, LDpred-p+t, LDpred-inf, LDpred2-Auto, LDpred2-Inf, PANPRS, DBSLMM, smtpred-wMtOLS, tlpSum, EB-PRS, PRSbils, and Plink. These methods are broadly accessible: they require modest input data, impose limited LD modelling burden, and ran reliably across phenotypes. Tools in the upper-right quadrant --- including PRSice-2, LDAK-GWAS, LDpred2-Grid, Lassosum, GEMMA-LMM, GEMMA-LM, SBayesRC, PRScs, and CTPR --- achieved strong predictive performance but at higher operational cost, primarily driven by reference panel requirements, full genotype-based LD estimation, runtime, or memory demand. Tools in the lower-left quadrant, including VIPRS-Simple, GCTA, SCT, HAIL, JAMPred, RapidoPGS-single, LDAK-GenotypeData, SDPR, and LDpred-gibbs, delivered below-median performance with low complexity and may be appropriate when computational resources are the primary constraint. Tools in the lower-right quadrant --- including BOLT-LMM, NPS, PleioPred, GEMMA-BSLMM, MTG2, AnnoPred, PRScsx, LDpred-funct, LDpred-fast, SBayesR, VIPRS-Grid, smtpred-wMtSBLUP, PolyPred, and CTSLEB --- incurred the highest operational burden without commensurate performance gains, suggesting they are best reserved for settings with specific methodological requirements rather than as default choices.

\subsection*{Tool failures reveal implementation and data-compatibility constraints}

A substantial number of PRS tools failed under at least one phenotype condition, as indicated by \texttt{NA} values in the full-model performance matrix (Table~\ref{TableBestFullModelSorted}). To investigate these cases systematically, we reviewed execution logs and classified failures into a structured taxonomy, including installation failure, dependency failure, input-format incompatibility, insufficient SNP overlap, invalid heritability estimate, phenotype-type incompatibility, reference-panel requirement not met, runtime/resource failure, and missingness/genotyping-rate constraint.

Most failures were attributable to software constraints or input-data requirements rather than to the biological properties of the phenotype itself. Methods such as NPS and CTPR were particularly sensitive to genotype completeness and exact SNP matching. Multi-trait methods such as smtpred-wMtSBLUP were affected by limited SNP overlap across traits. PolyPred and related methods depended on successful access to reference panels and stable heritability estimates, whereas tools such as AnnoPred were constrained by legacy software dependencies and external annotation requirements. Runtime and resource constraints also contributed to failure, particularly for computationally demanding tools such as PleioPred and LDpred-fast. These findings are provided in Supplementary Table 5 (Tool failure taxonomy).

\subsection*{Operational burden differs substantially across PRS tools}

In addition to predictive performance, we evaluated runtime, memory consumption, and phenotype-level completion rates across PRS tools.

When grouped by methodological family, multi-trait methods showed the lowest mean runtime (0.65~h), followed by penalised regression methods (0.81~h), whereas Bayesian methods had the highest mean runtime (1.95~h). Linear and linear mixed model approaches showed an intermediate mean runtime (1.48~h). In contrast, multi-trait methods had the highest average memory consumption (3.39~GB), followed by Bayesian methods (2.85~GB), whereas penalised and linear/LMM approaches had lower mean memory requirements (1.66~GB and 1.65~GB, respectively) (Table~\ref{tab:runtime_family_summary}).

\begin{table}[ht]
\centering
\small
\setlength{\tabcolsep}{5pt}
\begin{tabular}{|p{2.2cm}|p{1.8cm}|p{2.2cm}|p{2.3cm}|p{2.1cm}|p{2.3cm}|}
\hline
\textbf{Family} & \centering\textbf{Number of Tools} & \centering\textbf{Mean Runtime (h)} & \centering\textbf{Median Runtime (h)} & \centering\textbf{Mean Memory (GB)} & \centering\textbf{Median Memory (GB)} \tabularnewline
\hline
Penalised   & \centering 8  & \centering 0.81 & \centering 0.34 & \centering 1.66 & \centering 1.59 \tabularnewline
\hline
Bayesian    & \centering 20 & \centering 1.95 & \centering 0.46 & \centering 2.85 & \centering 2.23 \tabularnewline
\hline
Multi-trait & \centering 6  & \centering 0.65 & \centering 0.17 & \centering 3.39 & \centering 3.56 \tabularnewline
\hline
Linear/LMM  & \centering 12 & \centering 1.48 & \centering 0.46 & \centering 1.65 & \centering 1.10 \tabularnewline
\hline
\end{tabular}
\caption{Summary of runtime and memory usage across PRS tool families.}
\label{tab:runtime_family_summary}
\end{table}
At the individual-tool level, VIPRS-Simple was the most computationally efficient method, with an average runtime of 0.0025~h, followed by HAIL (0.0046~h) and C+T (0.0131~h). In contrast, CTPR was the most time-consuming method in the benchmark (18.67~h), followed by Lassosum (10.49~h) and PLINK (5.74~h). Memory usage also varied substantially, ranging from 0.26 GB for CTPR to 11.15 GB for AnnoPred, with GEMMA-LM (8.35 GB) and PRSice-2 (6.95 GB) also showing comparatively large memory footprints.

These results highlight that tool selection in applied PRS analyses depends not only on predictive performance, but also on computational feasibility and software robustness. In practical terms, methods with competitive predictive accuracy may differ substantially in runtime, memory requirements, and reproducibility under real analysis conditions. Detailed runtime and memory summaries are provided in Supplementary Table~2 (PRS tool characteristics).

\subsection*{Hyperparameter sensitivity and overfitting behaviour vary by method}

To evaluate whether predictive performance was influenced by model configuration, we performed a systematic hyperparameter sensitivity analysis across all PRS tools and phenotypes. For each tool, fold-level results were pooled across all phenotypes, and hyperparameters exhibiting variation were tested for association with held-out predictive performance using rank-based statistical tests. Hyperparameters were considered influential when variation in their values was significantly associated with changes in the predictive performance metric. Effect sizes were quantified using Spearman's correlation for numeric parameters and variance-based measures for categorical parameters, with uncertainty estimated using bootstrap-derived 95\% confidence intervals (CIs).

Across multiple tools, the GWAS p-value threshold emerged as one of the most consistently influential hyperparameters. In AnnoPred, the p-value threshold strongly influenced predictive accuracy for Height (effect size $= 0.72$, 95\% CI: $0.67$--$0.76$, $p = 2.8 \times 10^{-65}$), and similar patterns were observed for BOLT-LMM (effect size $= 0.74$, 95\% CI: $0.67$--$0.80$, $p = 4.7 \times 10^{-53}$) and C+T (effect size $= 0.70$, 95\% CI: $0.65$--$0.75$, $p = 1.16 \times 10^{-75}$), suggesting that the variant-inclusion threshold is a key determinant of predictive performance across multiple PRS construction strategies.

The number of variants included in the model was a second major driver of performance. In BOLT-LMM, the number of variants showed strong effects across several binary phenotypes, including Gastro-Reflux (effect size $= 0.84$, 95\% CI: $-0.88$ to $-0.76$, $p = 1.25 \times 10^{-22}$), High Cholesterol (effect size $= 0.87$, 95\% CI: $0.85$--$0.87$, $p = 4.9 \times 10^{-38}$), and Hypothyroidism (effect size $= 0.88$, 95\% CI: $0.86$--$0.89$, $p = 3.5 \times 10^{-38}$), indicating that the effective number of SNPs incorporated into the score substantially determines predictive performance in several tools.

Heritability-related parameters also influenced model performance in several Bayesian methods. In DBSLMM, the assumed trait heritability parameter was strongly associated with prediction accuracy for High Cholesterol (effect size $= 0.73$, 95\% CI: $-0.75$ to $-0.71$, $p = 4.8 \times 10^{-102}$) and IBS (effect size $= 0.71$, 95\% CI: $0.64$--$0.78$, $p = 1.53 \times 10^{-84}$), highlighting that the assumed genetic architecture encoded through heritability priors can substantially alter model behaviour. For mixed-model approaches, multiple parameters showed significant associations concurrently. In GCTA, the number of variants, heritability estimates, and regularisation parameters all demonstrated significant relationships with predictive performance across phenotypes, including a strong heritability--performance association for High Cholesterol (effect size $= 0.68$, 95\% CI: $-0.73$ to $-0.61$, $p = 3.0 \times 10^{-43}$). In GEMMA-LM, the p-value threshold strongly influenced performance for Height (effect size $= 0.67$, 95\% CI: $0.57$--$0.78$, $p = 1.39 \times 10^{-14}$) and IBS (effect size $= 0.53$, 95\% CI: $0.38$--$0.66$, $p = 1.95 \times 10^{-7}$).

In contrast, some tools exhibited limited sensitivity to hyperparameter variation. CTSLEB showed variation in p-value thresholds across phenotypes, but none reached statistical significance for predictive performance. GCTB-SBayesRC similarly showed hyperparameter variation without a statistically significant effect on prediction accuracy, suggesting that certain models are relatively robust to moderate changes in parameter settings.

Together, these results demonstrate that PRS predictive performance is strongly influenced by several key hyperparameters, particularly variant-inclusion thresholds, the number of variants used in model construction, and parameters governing trait heritability or genetic architecture, although the degree of sensitivity varies considerably between tools. This underscores the importance of systematic hyperparameter optimisation when benchmarking or deploying PRS models. Complete hyperparameter sensitivity results are provided in Supplementary Table~S6 (Hyperparameter sensitivity).

 \subsection*{Similarity of SNP effect-size profiles reflects methodological families}

To assess whether different PRS tools yielded similar SNP effect-size profiles, we compared the SNP effect-size estimates from each method across phenotypes. For each phenotype, tool-specific beta estimates were averaged across the five cross-validation folds, aligned on overlapping SNPs, and used to compute pairwise Pearson correlations between tools. These phenotype-specific correlation matrices were then averaged across phenotypes to obtain an overall cross-phenotype similarity matrix.

This analysis showed that several PRS tools produced closely related SNP effect-size profiles, whereas others formed more distinct groups. In general, methods with similar modelling strategies tended to cluster together. LDpred-based methods formed a coherent group, as did several tools relying on training genotype data for beta estimation. By contrast, some Bayesian and mixed-model methods showed weaker concordance with linear-model-based approaches, indicating that similar predictive performance can still arise from different underlying SNP effect-size architectures.

\begin{figure}[!ht]
\centering
\includegraphics[width=0.9\textwidth]{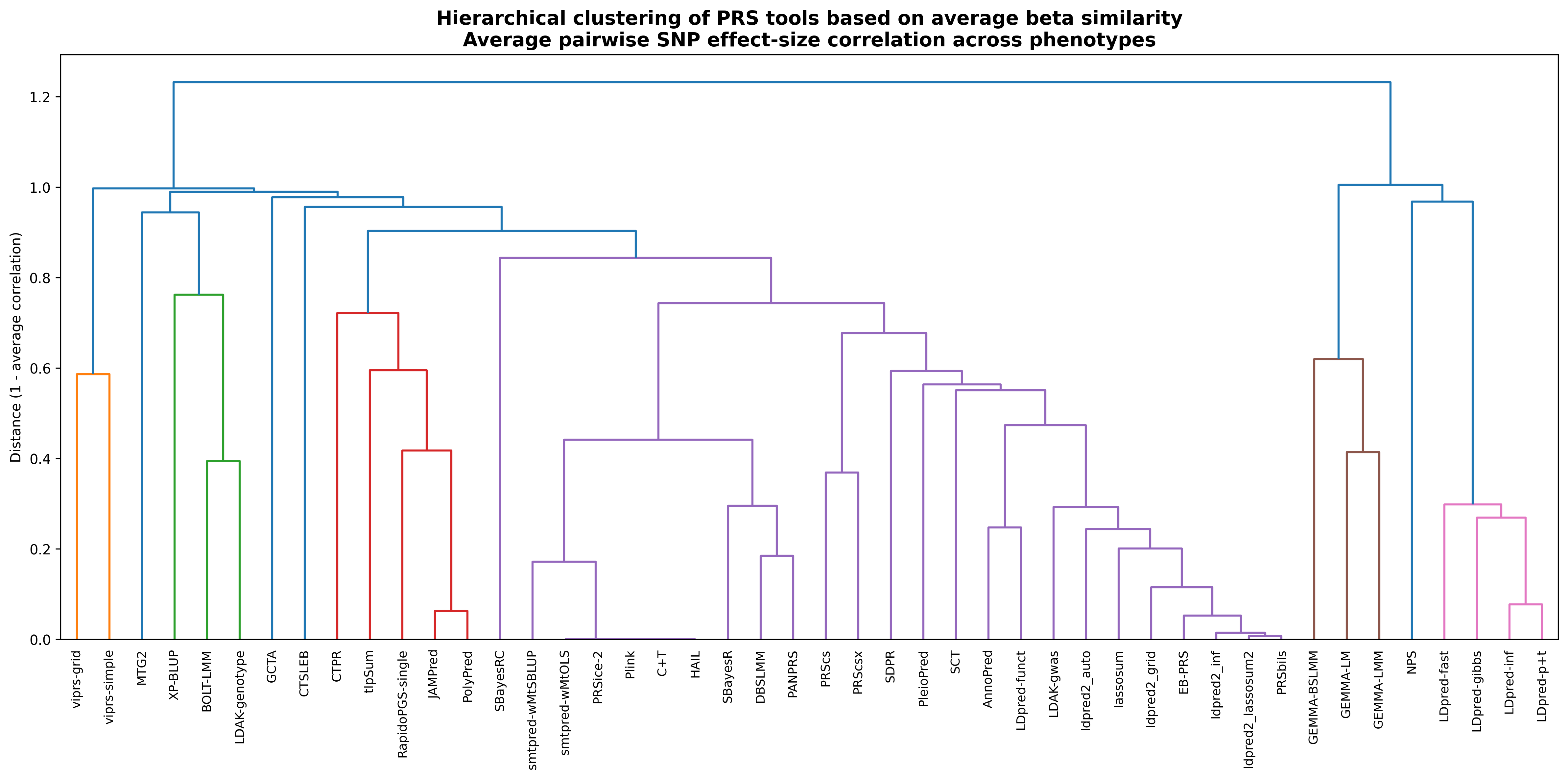}
\caption{Hierarchical clustering of PRS tools based on cross-phenotype similarity of SNP effect sizes. For each phenotype, tool-specific beta estimates were averaged across the five cross-validation folds and aligned on overlapping SNPs. Pairwise Pearson correlations between tools were then computed and averaged across phenotypes to obtain an overall similarity matrix. The dendrogram was constructed using distance defined as $1 - r$, where $r$ is the average pairwise correlation. Tools that cluster more closely produced more similar SNP effect-size profiles across the evaluated phenotypes.}
\label{fig:AverageBetaCorrelationDendrogram}
\end{figure}

The hierarchical dendrogram shows that PRS tools differ not only in predictive accuracy, but also in the structure of the effect sizes they generate. This distinction is useful for understanding methodological redundancy, complementarity, and the extent to which apparently similar tools are capturing comparable or divergent genetic signals. The corresponding visual summaries are shown in  Figure~\ref{fig:AverageBetaCorrelationDendrogram}. Additional phenotype-specific correlation summaries are provided in Supplementary Table~S7 (Beta's Correlation).

\section*{Discussion}

This study presents a harmonized benchmarking framework for the comparative evaluation of 46 PRS tools across multiple phenotypes. The framework standardizes data preprocessing, tool execution, hyperparameter exploration, and evaluation procedures, enabling a consistent comparison of diverse PRS methodologies under a common analytical pipeline. Our results demonstrate that PRS tool performance is strongly phenotype-dependent, with different tools performing optimally under different trait architectures and modelling conditions. This finding reinforces observations from previous studies that no single PRS method universally outperforms all others, highlighting the importance of evaluating multiple tools when constructing PRS models.

A central contribution of this work is the explicit separation of three model configurations --- null, PRS-only, and full --- across all tools and phenotypes. This design allows the predictive contribution of the PRS to be distinguished from that of the covariate structure, which is important because the binary phenotype models include an information-rich set of 135 NMR metabolomic biomarkers and comorbid conditions. By reporting all three configurations, the framework enables readers to evaluate PRS tools both in isolation and within a realistic prediction context, and to identify cases where apparent tool superiority reflects covariate interactions rather than differences in the underlying genetic modelling strategy. This transparency is a prerequisite for fair benchmarking in rich covariate settings and distinguishes the present framework from comparisons that report only a single final performance measure. It is also important to note that fold-level statistical comparisons between null and full models were limited by the five-fold design, which constrains the minimum attainable p-value. Conclusions about PRS utility are therefore grounded in effect size and directional consistency rather than in formal hypothesis rejection.

The systematic hyperparameter sensitivity analysis, conducted across all tools and phenotypes, revealed that variant-inclusion thresholds, number of variants, and heritability-related parameters are the most consistently influential drivers of predictive performance. This finding has practical implications for PRS deployment: it suggests that performance differences between tools may in some cases reflect differences in default hyperparameter settings rather than inherent methodological superiority. Benchmarking frameworks that do not explore the hyperparameter space risk producing rankings contingent on arbitrary default configurations. By pooling results across all phenotypes and folds, the present analysis provides a more robust characterisation of sensitivity than single-phenotype evaluations.

The cross-phenotype analysis of SNP effect-size profiles further showed that methodologically similar tools tend to produce correlated beta estimates, with LDpred-based methods forming a coherent cluster and genotype-dependent tools forming another. This clustering is informative for understanding methodological redundancy: tools that produce highly correlated effect-size profiles are unlikely to offer complementary predictive information even when their summary performance metrics differ. Conversely, tools that are statistically similar in predictive accuracy but divergent in their effect-size profiles may capture different aspects of the underlying genetic architecture, with implications for ensemble or stacking approaches.

Beyond predictive accuracy, the benchmarking framework captures operational characteristics that substantially influence the practical deployment of tools. Considerable variation was observed in runtime, memory consumption, and execution robustness. Some tools completed analyses within minutes, whereas others required many hours or substantial memory resources. Several methods failed under specific data conditions due to strict input requirements, dependence on reference panels, or unstable heritability estimates. These results reinforce that operational considerations --- including computational efficiency, installation complexity, and robustness to real-world data constraints --- are important selection criteria alongside predictive performance, particularly in multi-phenotype or large-scale genomic analyses. 

The sensitivity analysis of the hyperparameter selection rule provided an additional methodological insight beyond validating the $\delta$-constrained approach. When rankings were compared between the stability-constrained rule and a training-only rule that imposes no generalisation constraint, methods relying on full genotype-based LD modelling or linear mixed-model estimation --- including GEMMA-LMM, GEMMA-LM, and MTG2 --- showed the largest rank deterioration, indicating that their configurations are particularly susceptible to overfitting when selected without a stability constraint. By contrast, summary-statistics-based and clumping methods such as PRSice-2, Lassosum, and LDpred-p+t were robust across all selection conditions. This pattern suggests that the degree to which a PRS method relies on individual-level genotype data for model fitting is a predictor of its sensitivity to configuration selection strategy, with implications for how such methods should be deployed and validated in applied settings.

Several limitations should be acknowledged. First, the benchmarking was performed across a limited number of phenotypes and datasets drawn from a single-cohort setting, which may limit the generalisability of the observed performance patterns to other ancestries, phenotype definitions, or data ascertainment strategies. Second, the deeper analyses of SNP effect-size clustering and hyperparameter sensitivity were summarised across phenotypes rather than examined individually for each trait, which may obscure phenotype-specific patterns in method behaviour; individual phenotype-level results are provided in the supplementary material to allow further inspection. Third, differences in software requirements and input assumptions among tools make uniform benchmarking challenging despite the standardized execution framework, and some tools may be evaluated only on a subset of phenotypes due to software or data compatibility constraints.

Future work should extend the benchmarking framework to larger cohorts, more diverse ancestry groups, and a broader range of continuous and binary phenotypes to evaluate the robustness and transferability of the observed performance patterns. Integration of ancestry-aware evaluation and cross-cohort replication would be particularly valuable given the known sensitivity of PRS performance to ancestry match between training and target populations.

Overall, this framework provides a structured and reproducible platform for comparative evaluation of PRS tools. By integrating predictive performance metrics with operational characteristics such as runtime, memory usage, hyperparameter sensitivity, and software robustness, the framework provides a more comprehensive perspective on the utility of PRS tools than performance-only comparisons and is designed to support both applied tool selection and future methodological benchmarking.

To ensure full reproducibility, all analyses were executed using Conda-managed environments with explicitly defined software versions. Because several PRS tools require incompatible dependencies and different language versions, the workflow was organised into five dedicated environments: genetics (Python 3.10, R 4.3), ldscc (Python 2.7), advanceR (R 4.3), polyfun (Python 3.10), and viprs\_env (Python 3.10, R 4.3). Each environment is defined by a version-controlled environment.yml configuration file that specifies all package dependencies required for tool execution. Replication instructions and environment configuration files are provided with the framework to allow the complete computational setup to be recreated.

Several methods were evaluated but excluded from the final benchmarking due to practical limitations. Multiprs represents a methodological framework rather than a standalone tool, BGLR-R returned NA explained variance values across all phenotypes despite correct input formats, and PolyRiskScore is a web-based platform with limited flexibility for large-scale benchmarking. In addition, FairPRS does not support parallel execution across datasets on HPC systems, and RapidoPGS-multi does not support continuous phenotypes.

Taken together, this study provides a reproducible, implementation-aware benchmarking resource that captures both the predictive and operational dimensions of PRS tool performance. The proposed framework serves as a practical foundation for future comparative studies across broader cohorts, diverse ancestry groups, and additional phenotype architectures, and may assist researchers in selecting and evaluating PRS methods for applied genomic analyses.

\section*{Declarations}
 \backmatter

\subsection*{Supplementary information}
The article includes a supplementary file, \textbf{Supplementary Material 1.xlsx}, which contains additional tables supporting the benchmarking analysis.

Supplementary Table~S1 lists the phenotypes evaluated in this study together with the corresponding GWAS datasets used, including the GWAS Catalog accession identifiers or source dataset names.

Supplementary Table~S2 provides a consolidated overview of the PRS tools included in the benchmarking framework, including tool links, required input data, language, method family, computational characteristics, software environments, and summary benchmarking statistics.

Supplementary Table~S3 reports the predictive performance of each PRS tool across all evaluated phenotypes and cross-validation folds.

Supplementary Table~S4 summarizes the training and test performance for the null, PRS-only, and full models for each tool and phenotype, together with the gain obtained after inclusion of the PRS in the covariate-adjusted model.

Supplementary Table~S5 provides a structured taxonomy of PRS tool failures observed during benchmarking, including the affected phenotypes, failure stage, failure category, and observed cause.

Supplementary Table~S6 presents the statistically significant hyperparameters influencing PRS predictive performance. For each PRS tool and phenotype, hyperparameters that are significantly associated with predictive performance are reported. Effect sizes are Spearman correlations between hyperparameter values and held-out model performance (\texttt{Test\_best\_model}), and confidence intervals are bootstrap-derived 95\% confidence intervals.

Supplementary Table~S7 reports the cross-tool correlation of SNP effect sizes (beta estimates). For each phenotype, beta estimates were averaged across the five cross-validation folds, aligned on overlapping SNPs, and used to calculate pairwise correlations between tools. These phenotype-specific correlations were then summarized across phenotypes to provide an overall measure of similarity between PRS methods.

\subsection*{Competing interests}
The authors declare no competing interests.

\subsection*{Ethics approval and consent to participate}
Not applicable

\subsection*{Data availability }
GWAS summary statistics for the binary phenotypes were downloaded from the GWAS Catalog (\url{https://www.ebi.ac.uk/gwas/}). The genotype data were accessed from the UK Biobank under application ID 50000 (\url{https://www.ukbiobank.ac.uk/}). For Height, we used the GWAS and genotype data from the PRS tutorial dataset described by Choi et al.~\cite{Choi2020}, which can be downloaded from \url{https://choishingwan.github.io/PRS-Tutorial/base/}.

\subsection*{Code Availability}
The code developed for this research is available on GitHub at \url{https://github.com/MuhammadMuneeb007/PRSTools}, and the documentation is available at \url{https://muhammadmuneeb007.github.io/PRSTools/Introduction.html}.

\subsection*{Author contribution}
M.M. implemented the code segments, produced the online documentation, and wrote the initial draft of the manuscript. D.A. supervised the research and revised and edited the methodology and manuscript.

\bibliography{sn-bibliography}

\end{document}